\documentclass[aps, pra, superscriptaddress, showpacs, floatfix, longbibliography, twocolumn]{revtex4-1}

\usepackage{amsmath}
\usepackage{amssymb}
\usepackage{bbold}
\usepackage{graphicx}
\usepackage{epstopdf}
\usepackage{color}
\usepackage{soul}
\usepackage{hyperref}
\usepackage{bm}

\hypersetup{
  colorlinks = true, 
  urlcolor   = blue, 
  linkcolor  = blue, 
  citecolor  = blue  
}

\begin{document}

\title{Phase diagram of a pseudogap Anderson model with application to graphene}
\author{Hung T. Dang}
\affiliation{Faculty of Materials Science and Engineering, Phenikaa University, Yen Nghia, Ha Dong district, Hanoi, Vietnam}
\affiliation{Phenikaa Institute for Advanced Study, Phenikaa University, Yen Nghia, Ha Dong district, Hanoi, Vietnam}
\author{Hoa T.M. Nghiem}
\affiliation{Phenikaa Institute for Advanced Study, Phenikaa University, Yen Nghia, Ha Dong district, Hanoi, Vietnam}
\date{\today}

\begin{abstract}
The Anderson model of an $s$-wave single-orbital correlated impurity placed on a noninteracting honeycomb lattice, a simplified model for studying an impurity on graphene, is used to investigate pseudogap Kondo problem. In this model, there are two quantum phases: the phase of free impurity local moment and the Kondo phase where this local moment is fully screened. The transition between these two phases is under investigation. The work focuses mostly on the case where the impurity is placed on top of a lattice site. In this case, the full phase diagram is constructed using three parameters: the Hubbard interaction $U$, the hybridization strength $v_0$ and the impurity energy level $\epsilon_d$. The phase diagram exhibits linear $(U^c, \epsilon_d^c)$ phase boundary, the slope of which, as well as the critical value $\epsilon_d^c$, depends strongly on $v_0^2$. Further analysis shows that the real part of the self energy at zero frequency and the impurity occupancy can help to understand the behaviors of the phase boundaries. The dependence of the phase transition on the impurity position is briefly discussed, revealing difficulties that one needs to solve in order to realize the pseudogap Kondo model in the realistic graphene lattice.
\end{abstract}

\maketitle

\section{Introduction \label{sec:intro}}

The Kondo effect, a correlation effect where the spin of an impurity is completely screened by conduction electrons of the host material at low temperature, is one of the most fundamental problems in the field of strongly correlated systems \cite{Hewson1997}. Since the theoretical work of Kondo studying the anomalous resistivity of metallic systems doped with magnetic impurities \cite{Kondo1964} in the 1960s, the Kondo model has revealed interesting universal behaviors at temperatures below a Kondo energy scale $T_K$ related to the suppression of the impurity local moment \cite{Wilson1975}. The Kondo model and the more general Anderson model \cite{Anderson1961} are considered as the most basic models to study many-body physics, which is rather well-understood especially with the invention of the numerical renormalization group (NRG) method in 1970s \cite{Wilson1975,Bulla2008}. The Kondo physics plays an important role in understanding more complicated strongly correlated problems such as the Kondo lattice model for heavy-fermion systems \cite{Doniach1977,Loehneysen2007,Gegenwart2008}, lattice correlated systems such as the Hubbard model or multi-orbital systems \cite{Georges1992,Georges1996,Georges2013}. From the methodology point of view, solving the correlated impurity problem allows scientists to further develop more powerful methods for correlated lattice systems, in particular the dynamical mean-field theory \cite{Georges1996,Gull2011}.

Since 1990s, the topic of Kondo problems has been ``revived'' \cite{Kouwenhoven2001} thanks to new experimental realizations of Kondo systems in quantum dots \cite{GoldhaberGordon1998,Sasaki2000} and by controlling magnetic adatoms on metallic surfaces using scanning tunneling microscopy \cite{Crommie1993a,Crommie1993b,Li1998,Manoharan2000}. Moreover, there are corners in the field of Kondo problems that are not thoroughly investigated and are still challenging to solve theoretically or to measure experimentally. One of them is the class of impurity problems where the host materials are semimetals, of which the low-energy density of states (DOS) vanishes at the Fermi level as a power law of frequency $\nu(\omega) \sim |\omega|^r$. It is called the ``pseudogap'' Kondo problem \cite{Withoff1990,Gonzalez-Buxton1998}. The vanishing point at the Fermi level of the DOS of the host material may prevent the Kondo screening effect to occur even at nonzero coupling constant, thus there exists a quantum phase transition from the free local moment phase to the Kondo phase, where this local moment is fully screend, by varying the coupling constant or by increasing valence mixing \cite{Withoff1990,Pixley2012}. This is different from the conventional Kondo system (the DOS of the host is nonzero at the Fermi level), where the state of the free local moment only appears when the Kondo coupling vanishes. Previous works \cite{Withoff1990,Bulla1997,Gonzalez-Buxton1998} have shown that different quantum phases occur depending on the value of $r$, which is divided into three regions: small $r$ ($0<r<r^*$), intermediate $r$ ($r^* < r < r_{max}$) and large $r$ ($r > r_{max}$) with $r^*\approx 0.375$ and $r_{max}=\frac{1}{2}$. Small- and intermediate-$r$ regions are distinguished from the large-$r$ one by the quantum critical point (QCP) at the particle-hole symmetric point, which is absent in the latter case. The intermediate-$r$ region is differentiated from the small-$r$ region by an additional asymmetric fixed point when the particle-hole symmetry is broken \cite{Gonzalez-Buxton1998}. In this region, $r=1$ is the case of interest. Cassanello and Fradkin \cite{Cassanello1996} has shown that, at $r=1$, there are logarithmic corrections included in the scaling laws of many kinds of physical observables, suggesting that the physics at $r=1$ be analogous to a system at an upper critical dimension at criticality. Further study \cite{Vojta2004,Fritz2004} provides more evidence for the characteristics of upper critical dimension at $r=1$ such as the nature of the phase transition is changed when crossing $r=1$ and perturbative approach can be feasible at $r > 1$.

In the experimental aspect, the pseudogap Kondo model has been used to interpret reasonably a new type of QCPs that occurs in heavy fermion systems \cite{Si2001,Gegenwart2008,Pixley2012} or to study impurity local moment in unconventional superconductors \cite{Vojta2001}. However, the difficulty in realizing the model in experiments is the main issue that hinders the progress of research in this topic. Indeed there are not many materials with power-law DOS around the Fermi level, a few candidates are $d$-wave superconductors and graphene which are related to the case $r=1$. Studying impurities in $d$-wave superconductors is nontrivial because directly measuring properties at the impurity in the bulk system is not simple. Graphene on the other hand might be more prospective experimentally as it is a 2D material, allowing for surface measurement techniques such as scanning tunneling microscopy. Not so long after its discovery \cite{Novoselov2004}, graphene has been used as the host material for magnetic adatoms, typically the cobalt adatoms \cite{Mattos2009,Jacob2009,Brar2011,Wang2012}. Much effort is devoted to the determination of the stable position of an adatom on graphene and the dependence of its magnetic state on each position \cite{Wehling2010a,Wehling2010b,Eelbo2013,Virgus2014}. Furthermore, for most of the cases, graphene is deposited on a substrate, the substrate is thus another factor that strongly affects the magnetic behavior of the adatom on graphene \cite{Ren2014,Donati2014}. The readers may also refer to Ref.~\cite{Fritz2013} for a review of theories and experiments for the problem of an impurity on graphene lattice.

This paper is devoted to the understanding of the pseudogap Kondo problem where the host metal, a honeycomb lattice, simulating the low-energy physics of graphene, and the adatom is restricted to an $s$-wave single-orbital impurity. From our high-temperature simulation data generated by continuous-time quantum Monte Carlo method \cite{Gull2011}, we employ Binder analysis \cite{Binder1981,Pixley2012} to determine the phase transition at zero temperature and study physical properties as the temperature decreases. We focus mostly on the case of an impurity placed on top of a lattice site. First we build a full phase diagram for the $r=1$ pseudogap Anderson model, in which the Kondo screening phase can only be observed in a mixed-valent state by driving the impurity away from half-filling. We then discuss briefly the situations where the impurity is placed at different location on the lattice. Interpretation of the phase diagram and discussion of different impurity positions on the lattice give us a broader view about the possibility for the realization of the pseudogap Kondo model on graphene.

The rest of the paper is structured as follows. Section~\ref{sec:model_methods} describes the model and basic parameters of the system under investigation, as well as the simulation method in use (mainly the continuous-time quantum Monte Carlo method \cite{Gull2011}). Section~\ref{sec:t_pos} presents the full phase diagram of the pseudogap Anderson model when the impurity is placed on top of an atom in the honeycomb lattice. Section~\ref{sec:analyze} analyzes the behaviors of each phase boundary presented in Sec.~\ref{sec:t_pos}. Section~\ref{sec:bh_pos} discusses the situations where the impurity is placed at different locations in the host material. Finally, Section~\ref{sec:conclusions} concludes our study.

\section{Model and Methods \label{sec:model_methods}}

\subsection{Model}

\begin{figure}[t]
 \includegraphics[width=\columnwidth]{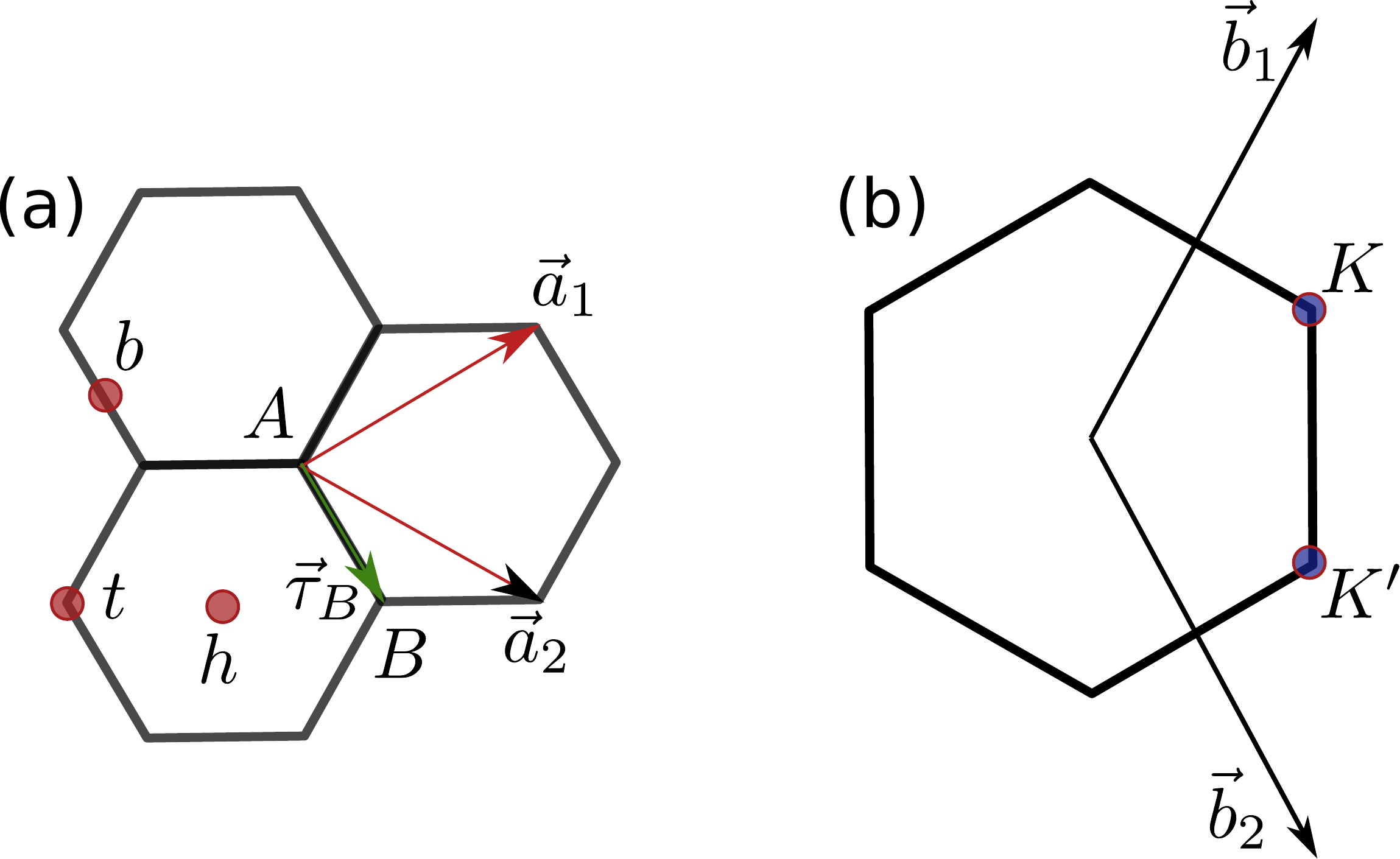}
\caption{\label{fig:imp_graphene} (a) Three possible possitions of an impurity on the honeycomb lattice (graphene): at the hollow site ($h$), at the bridge site ($b$), and at the top site ($t$). The honeycomb lattice vectors are $\bm{a}_1 = (3/2, \sqrt{3}/2)$, $\bm{a}_2 = (3/2, -\sqrt{3}/2)$ where the length of each edge of the hexagon is set to 1. A unit cell has two atoms $A$ and $B$ at positions $\bm{\tau}_A = (0, 0)$ and $\bm{\tau_B} = (1/2, -\sqrt{3}/2)$ corresponding to the two sublattices. (b) The first Brillouin zone (BZ1) of this honeycomb lattice is defined by two reciprocal vectors $\bm{b}_1$ and $\bm{b}_2$. The two valleys are at $K=(2\pi/3, 2\pi/(3\sqrt{3}))$ and $K'=(2\pi/3, -2\pi/(3\sqrt{3}))$.}
\end{figure}

We consider the problem of a correlated impurity placed on an uncorrelated metallic layer. The host metallic layer is a honeycomb lattice with only one uncorrelated orbital per site, simulating a graphene sheet [see Figure~\ref{fig:imp_graphene}(a)]. The impurity is assumed to have only one correlated $s$ orbital. Fig.~\ref{fig:imp_graphene}(a) marks three impurity positions of high probability on graphene \cite{Wehling2010a,Eelbo2013,Virgus2014} that we take into account: (a) impurity located on top of an atom (the top site $t$), (b) at the center of the hexagon of six atoms (the hollow site $h$), (c) in the middle of two neighboring atoms (the bridge site $b$).

We focus on a more general Anderson impurity model \cite{Anderson1961}, of which the Kondo model is a special case. The Hamiltonian has three parts: the Hamiltonian of the host material ($H_{host}$), the local impurity Hamiltonian ($H_{loc}$), and the hybridization between the impurity and the host ($H_{hyb}$).
\begin{equation}
\label{eq:hamiltonian}
 H = H_{bath} + H_{loc} + H_{hyb}.
\end{equation}

The Hamiltonian of the host material is constructed using the tight binding approach, where conduction electrons can only hop from one site to sites nearby. The nearest neighbor hopping $t$ is the most important one, and is used as the basic energy scale in this paper. The next-nearest neighbor conduction electron hopping $t'$, if there is, only plays the role of an additional chemical potential, and thus is neglected in this study. Thus the analytic form, which can be obtained elsewhere \cite{Neto2009,Fritz2013}, reads
\begin{equation}
\label{eq:h_bath}
 H_{bath} = \sum_{\bm{k}\sigma}
 \begin{pmatrix}
  c^\dagger_{A\bm{k}\sigma} & c^\dagger_{B\bm{k}\sigma}
 \end{pmatrix}
 \cdot
 \hat{K}_{\bm{k}}
 \cdot
 \begin{pmatrix}
  c_{A\bm{k}\sigma} \\
  c_{B\bm{k}\sigma}
 \end{pmatrix},
\end{equation}
where
\begin{align}
\label{eq:tbham}
 \hat{K}_{\bm{k}} &= \begin{pmatrix}
  -\mu & \pi_{\bm{k}} \\
  \pi^*_{\bm{k}}  & -\mu
 \end{pmatrix},\\
 \pi_{\bm{k}} &= -t e^{i\bm{k}\bm{\tau}_B}[1 + e^{-i\bm{k}\bm{a}_2} + e^{i\bm{k}(\bm{a}_1-\bm{a}_2)}].
\end{align}
The operator $c_{\alpha \bm{k}\sigma}$ ($c^\dagger_{\alpha \bm{k}\sigma}$) is the annihilation (creation) operator of conduction electrons from the host, with $\alpha = A, B$ denoting the $A$ and $B$ sublattices of the honeycomb lattice [see Fig.~\ref{fig:imp_graphene}(a)], $\mu$ is the chemical potential. The position vector $\tau_A$ ($\tau_B$), specifying the location atom $A$ ($B$) in the unit cell, is defined in the caption of Fig.~\ref{fig:imp_graphene} and illustrated in the same figure. The $\bm{k}$ sum is carried out over the first Brillouin zone (BZ1) as plotted Fig.~\ref{fig:imp_graphene}(b).

It is known that the energy dispersion of graphene has two valleys $K$ and $K'$ in the BZ1 [see Fig.~\ref{fig:imp_graphene}(b)], expanding $k$ around those valleys results in linear dispersion $\epsilon_k \approx 3t' \pm v_F |k|$ where $v_F = \frac{3t}{2}$ is the Fermi velocity \cite{Neto2009}. Consequently, with $t'=0$, the DOS for this two-dimensional lattice is linear $\nu(\epsilon) \sim |\epsilon| / v^2_F$, which is the linear pseudogap feature required by our study. The chemical potential is set to $0$ to ensure that the Fermi level is at the charge neutral point, which is also necessary for the pseudogap Kondo model.

The local Hamiltonian of the impurity contains the onsite Hubbard interaction term with the correlation strength $U$ and the energy level $\epsilon_d$ of impurity electrons
\begin{equation}
\label{eq:h_loc}
 H_{loc} = \sum_{\sigma} \epsilon_{d} d^\dagger_{\sigma} d_{\sigma} + U n_{d\uparrow} n_{d\downarrow},
\end{equation}

Finally, the hybridization part reads
\begin{equation}
\label{eq:h_hyb}
 H_{hyb} = \sum_{\bm{k}\alpha\sigma} (V_{\alpha \bm{k}} c^\dagger_{\alpha \bm{k} \sigma} d_{\sigma} + h.c.).
\end{equation}
where $V_{\alpha \bm{k}}$ is the hybridization coefficient in the reciprocal space, $\alpha=A, B$ is the sublattice index, $\sigma$ is the spin index. Based on the hybridization in the real space $V(\bm{R}+\tau_\alpha, \bm{R}_{imp})$, which is usually limited to the electron hopping between the impurity and neighbor lattice sites, $V_{\alpha \bm{k}}$ is calculated as
\begin{equation}\label{eq:hybk}
 V_{\alpha \bm{k}} = \dfrac{1}{\sqrt{N}} \sum_R e^{-i\bm{k}(\bm{R}+\bm{\tau}_\alpha - \bm{R}_{imp})} V(\bm{R}+\bm{\tau}_\alpha, \bm{R}_{imp}).
\end{equation}
where $\bm{R}_{imp}$ is the position of the impurity, $\bm{\tau}_\alpha$ is the position of an atom in the unit cell, as defined above. Similar to the tight binding calculation for the host material, the dominant $V(\bm{R}+\bm{\tau}_\alpha, \bm{R}_{imp})$ terms in Eq.~\eqref{eq:hybk} that contribute 
to $V_{\alpha \bm{k}}$  are the nearest neighbor hybridizations. We denote these hybridizations as $v_0 = V(\bm{r}_{nn}, \bm{R}_{imp})$ where $\bm{r}_{nn}$ are the positions of the nearest-neighbor sites, which depend on the location of the impurity on the lattice. Hybridization to sites of larger distance are neglected. In our work, we consider mostly the hybridization in the dynamical form
\begin{equation}
\label{eq:delta}
\begin{split}
 \Delta_\sigma(z) &= \sum_k (V^*_{Ak}, V^*_{Bk}) \cdot (\mathbb{1} z-K_k)^{-1} \cdot 
 \begin{pmatrix}
  V_{Ak} \\ V_{Bk}
 \end{pmatrix},
\end{split}
\end{equation}
where $\mathbb{1}$ is the $2\times 2$ identity matrix. The variable $z$ represents general frequency. To obtain the Matsubara (real frequency) hybridization function, we replace $z$ by the Matsubara frequency $i\omega_n$ ($\omega + i 0^+$). The impurity Green function reads
\begin{equation}
\label{eq:Gimp}
 G_\sigma^{imp}(z) = \left[z - \epsilon_{d} - \Delta_\sigma(z) - \Sigma_\sigma(z)\right]^{-1},
\end{equation}
where the self-energy $\Sigma_\sigma^{imp}(z)$ characterizes the correlation effects of impurity electrons.

\subsection{Methods\label{subsec:method}}

For the entire study, the most difficult part is the treatment of the correlation effects emerging from the onsite Hubbard interaction of the impurity electrons. In this work, we employ the hybridization expansion version of the continuous-time QMC impurity solver \cite{Werner2006,Gull2011} (CT-HYB) provided by the TRIQS package \cite{Parcollet2015,Seth2016}. CT-HYB is our method of choice for several reasons: it provides a full treatment for the correlation with numerically exact solution; the simulation works for a wide range of input parameters and for any given DOS of the host material; finally, it allows us to measure dynamical quantities such as spin-spin correlation or the Green function of the impurity even at the critical point.

To ensure the quality of our CT-QMC calculations, we choose the technical parameters for the simulations as follows. The number of Monte Carlo updates between two consecutive measurements is set at $200$. Global update by swapping spin up and spin down impurity electron operators is conducted to avoid unrealistic spin polarization, and is proposed at the rate of $0.15$. For determining the phase transition, we require simulations with the total number of measurements per simulation of at least $5\times 10^6$ (typical number for large $v_0$ measurement ($v_0 > t$), for small $v_0$ it can reach as larger as $10^9$ measurements for each simulation). For the range of parameters under investigation, with $\beta$ ranging from $100/t$ to $400/t$, it takes $2000$ to $3500$ seconds to finish one calculation with $128$ CPU cores.

The biggest limitation in this method is that it is computationally expensive. Even without the sign problem \cite{Troyer2005}, the complexity increases as rapidly as $O\left(\frac{1}{T^3}\right)$ when the temperature decreases \cite{Gull2011}. Therefore, within this method, we are unable to reach low temperature regime well below the Kondo scale. However, by assuming the inverse temperature $\beta=1/T$ as the ``size'' of the system, one can apply techniques from large-scale Monte Carlo simulation for this study \cite{Glossop2011,Pixley2012,Binder1981}. Binder analysis \cite{Binder1981}, an effective method for studying critical point in large-scale Monte Carlo method, is employed to determine the QCPs. Following Refs.~\cite{Glossop2011,Pixley2012}, we set the Binder parameter as $B_T(\epsilon_d, v_0, U) = \dfrac{\langle m_z^4\rangle}{\langle m_z^2\rangle^2}$ where $m_z = n_{d\uparrow} - n_{d\downarrow}$.

\begin{figure}[t]
\centering
 \includegraphics[width=\columnwidth]{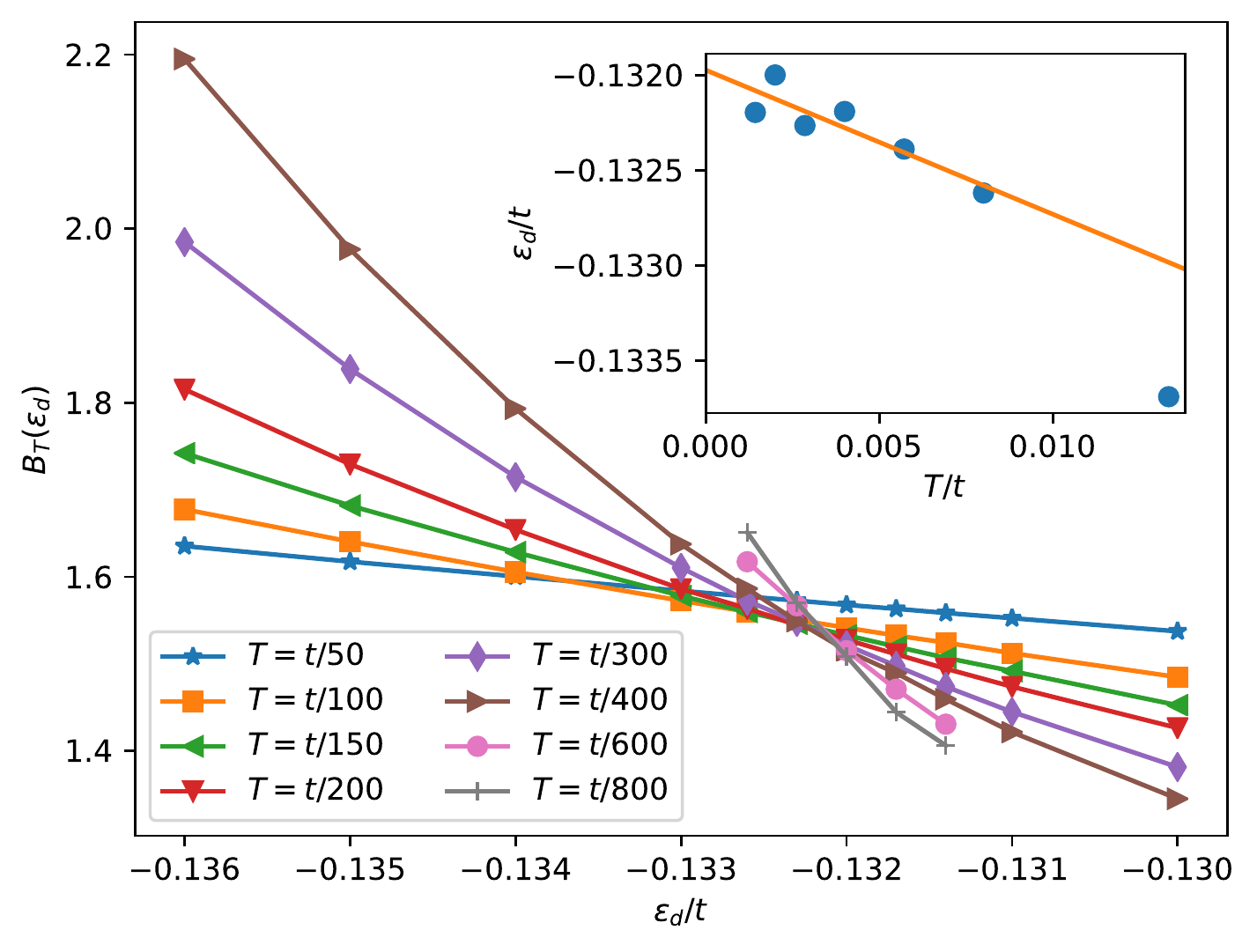}
\caption{\label{fig:benchmark} Demonstration for determining the critical point ($\epsilon_d^c$) at fixed $U=0.15t$ and $v_0=0.5t$: the Binder parameter $B_T$ vs. impurity energy level $\epsilon_d$ for several different temperatures $T$. The crossing point of two curves of two consecutive temperatures gives an estimate of $\epsilon_d^c$. Inset: the plot of $\epsilon_d(T)$ vs. $T$ for all crossing points in the Binder plot; the straight line is the linear fit for all but the point of the largest temperature ($T\approx0.0133t$). The intercept of the linear fit is the critical value $\epsilon_d^c = -0.13197(9)t$.}
\end{figure}

As demonstrated in Figure~\ref{fig:benchmark}, for all our calculations, we keep $v_0$ and $U$ fixed while varying $\epsilon_d$ and run simulations for several temperatures in order to determine QCPs. As a result, there are several $\epsilon_d$-dependent Binder curves, each of them corresponds to a specific temperature $T$. They nearly intersect at one point. The crossing point of each pair of curves of two consecutive temperatures $T_1$ and $T_2$ gives an estimate of the critical value $\epsilon_d^c(T)$ where $\frac{1}{T} = \frac{1/T_1 + 1/T_2}{2}$. By linear extrapolation to zero temperature, we obtain the critical $\epsilon_d^c$ [see the inset of Fig.~\ref{fig:benchmark}]. We note that, the linearity of the crossing points is not maintained for the whole range of temperature, e.g. in Fig.~\ref{fig:benchmark}, the crossing point of the highest temperature in the simulation $T\approx 0.0133 t$ stays out of the fitting line for the remaining points. Indeed, in most of our simulations, this temperature is out of the linear region for fitting, thus it is neglected in the extrapolation. On the other hand, simulations at low temperature require much longer time for reliable results, and a finer $\epsilon_d$ grid is needed for the linear extrapolation. For example, in the inset of Fig.~\ref{fig:benchmark}, the two points corresponding to the two lowest temperatures are computationally more expensive to obtain but do not significantly improve the extrapolation. Therefore, in this work, we choose the inverse temperatures $\beta = \frac{1}{T}$ in the range from $100/t$ to $400/t$ for linear extrapolating critical point.

Binder analysis does not only give us the position of the critical point $\epsilon_d^c$ but also help to distinguish phases. Considering the region below half-filling, for $\epsilon_d < \epsilon_d^c$, the Binder parameter becomes larger as $T$ decreases, signifiying the strong coupling between impurity local moment and conduction electrons, suggesting the Kondo screening phase. In contrast, for $\epsilon_d > \epsilon_d^c$, the Binder parameter decreases to $1$ as $T$ decreases, which is the signature that the local moment is decoupled and becomes a free moment, thus suggesting the free local moment phase \cite{Glossop2011,Pixley2012}. Fig.~\ref{fig:benchmark} illustrates these behaviors.

In this work, we vary all physical parameters for simulations to construct the full phase diagram. The set of parameters for investigation is $\{U,\epsilon_d, v_0\}$. $U$ is varied between $0$ and $0.2t$, $v_0$ is from $0$ to less than $2t$. The impurity energy level (plus the shifting due to the real part of the hybridization function if existing) is varied within the range $(-U, 0)$. The lowest temperature in use for constructing the phase diagram is $T = \frac{t}{400}$. Dynamical quantities such as the imaginary-time Green function $G(\tau)$, the self-energy $\Sigma(i\omega_n)$ and the spin-spin correlation function $C_s(\tau) = \langle \sigma_z(\tau) \sigma_z(0)\rangle$ (with $\sigma_z = n_{d\uparrow} - n_{d\downarrow}$) are also measured when necessary.

\section{Impurity on top of an atom \label{sec:t_pos}}

The main part of our study is the investigation for the case of an impurity on top of an atom in the honeycomb lattice (the $t$ site), which will be presented thoroughly in this and the next sections. We start by constructing the full phase diagram for this case. If considering the nearest neighbor hopping $t$ of conduction electrons as the basic energy scale and neglecting further hopping or hybridization, the phase diagram is a three-dimensional one with respect to the impurity energy level $\epsilon_d$, the Hubbard interaction strength $U$ and the nearest neighbor hybridization $v_0$.

Assuming the impurity is on top of an $A$ site in the honeycomb lattice, the hybridization function for the impurity is calculated using Eq.~\eqref{eq:delta} with $V_{Ak} = v_0$ and $V_{Bk} = 0$. It is indeed a replication of the corresponding honeycomb lattice Green function, differed only by a prefactor $\frac{v_0^2}{2}$. Thus the hybridization spectrum is related to the DOS of the lattice as
\begin{equation}\label{eq:t_hyb}
 -\dfrac{1}{\pi}\mathrm{Im}\Delta_\sigma(\omega) = \dfrac{v_0^2}{2} \nu(\omega).
\end{equation}
As we later examine in Fig.~\ref{fig:hyb_dos} the hybridization function for three impurity positions, the impurity at the $t$ site is the only case which exhibits symmetric hybridization spectrum. Thus it is the closest to the low-energy $r=1$-pseudogap Kondo model \cite{Gonzalez-Buxton1998,Pixley2012}. Therefore, in order to avoid unnecessary complication, the case of an impurity on top of a lattice site is under full consideration in this work.

\begin{figure}
 \includegraphics[width=\columnwidth]{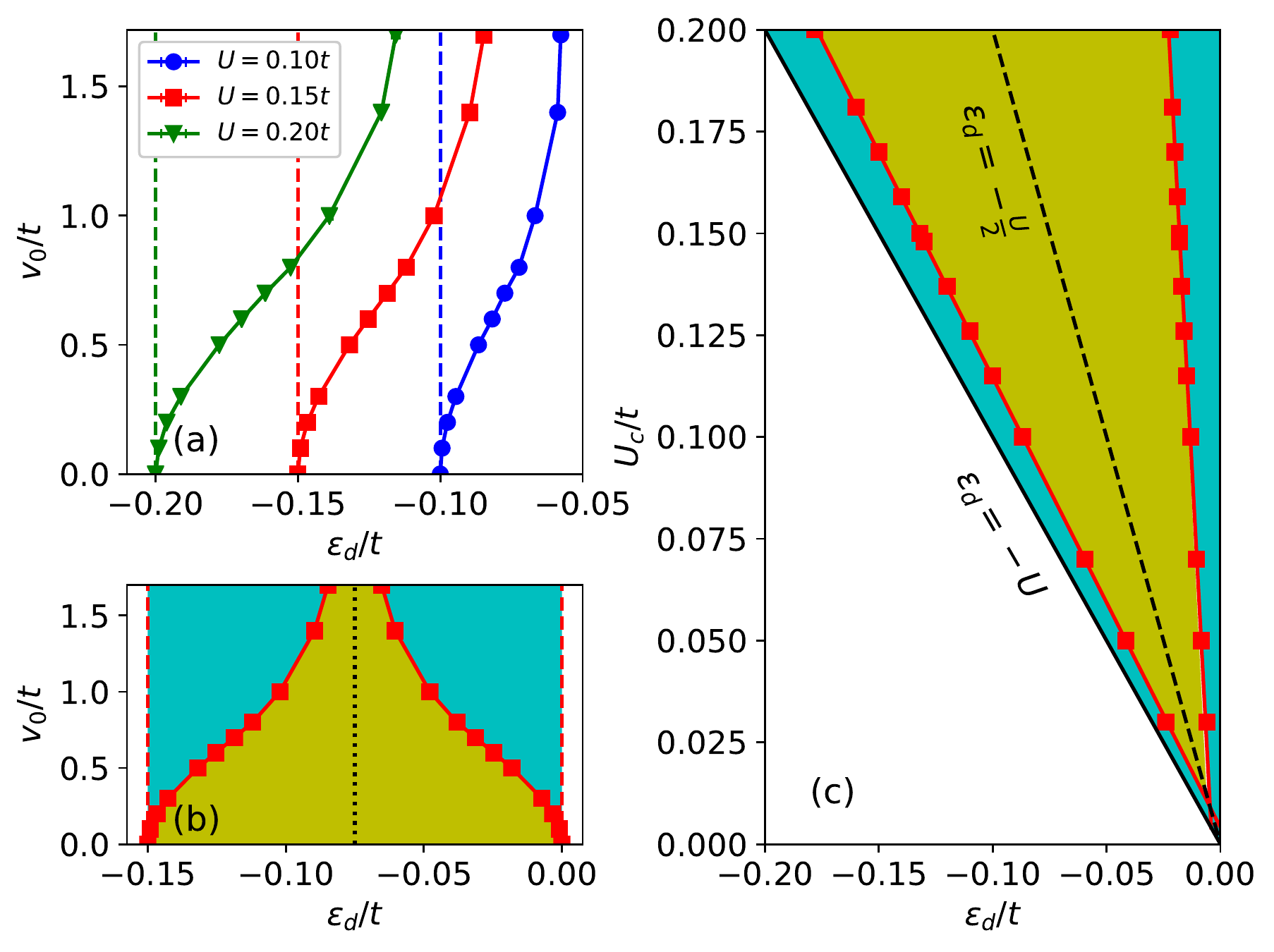}
\caption{\label{fig:phase_diagram} (a) The $v_0$ vs. $\epsilon_d$ phase boundaries of the pseudogap Anderson model of an impurity at the $t$ site for three different $U$ values (solid lines). The region bounded by a phase boundary and the corresponding $U = -\epsilon_d$ line (the vertical dashed line) is the Kondo screening phase, the region on the right of the phase boundary is the free local moment phase. (b) The full $v_0$ vs. $\epsilon_d$ phase diagram at $U=0.15t$. The vertical red dashed lines are $\epsilon_d=-U$ and $\epsilon_d=0$; the dotted line is $\epsilon_d = -\frac{U}{2}$. (c) The $U$ vs. $\epsilon_d$ phase diagram at $v_0=0.5 t$. In panels (b) and (c), the yellow regions mark the free local moment phase, the cyan regions mark the Kondo screening phase. The error bars the of points constructing the phase boundaries are smaller than the symbol size.}
\end{figure}

We present in Figure~\ref{fig:phase_diagram} the phase diagrams in the $(v_0, \epsilon_d)$ plane [panels (a) and (b)] and in the $(U, \epsilon_d)$ plane [panel (c)] where the remaining parameter is kept constant. As noted previously, because these are quantum phase transitions, it requires low temperature for investigation. For nonzero-temperature methods such as CT-HYB, we can employ Binder analysis (see Sec.~\ref{subsec:method}) to extrapolate critical points from high-temperature calculations \cite{Binder1981,Glossop2011,Pixley2012}. All the phase boundaries in Fig.~\ref{fig:phase_diagram} are determined in this way. To justify the Binder analysis approach, we present in Appendix~\ref{app:occ_susc} additional measurements of the occupancy and the spin susceptibility for a specific case to confirm the existence of the critical points.

For the pseudogap model where the DOS of the host material is $\nu(\omega) \sim |\omega|^r$ with $r > r_{max} = 1/2$, there is no phase transition at the particle-hole symmetric point $\epsilon_d = -\frac{U}{2}$, regardless of how large $U$ is \cite{Gonzalez-Buxton1998}. It means that at half-filling, the impurity local moment is always decoupled from the conduction electrons, i.e. the free local moment phase. Only when $\epsilon_d$ is shifted away from the particle-hole symmetry, a mixed-valence quantum critical point may occur \cite{Pixley2012} for the transition to the phase of screened local moment (the Kondo phase), as illustrated in Fig.~\ref{fig:phase_diagram}. In all phase diagrams, the line $\epsilon_d = -\frac{U}{2}$ divides the phase diagram into two regions above and below half-filling, each of which contains only one phase boundary line separating the two phases. For the region of less than half-filling ($\epsilon_d < -\frac{U}{2}$), the left side of the phase boundary is the Kondo screening phase, while the right side is the free local moment phase. Because of the particle-hole symmetry at $\epsilon_d = -\frac{U}{2}$, the phase diagram is symmetric about this line. Thus, in Fig.~\ref{fig:phase_diagram}(b) and (c), we plot the region of $\epsilon_d > -\frac{U}{2}$ as the mirror image of the other half-plane.

Focusing only on the region of less than half-filling $\epsilon_d < -\frac{U}{2}$, we find that the critical value $\epsilon_d^c$ is always in the range $[-U, -\frac{U}{2}]$ and it depends monotonically on $v_0$. In Fig.~\ref{fig:phase_diagram}(a) and (b), $\epsilon_d^c$ approaches $-U$ as $v_0 \to 0$ and tends to approach $-\frac{U}{2}$ as $v_0 \to \infty$. This dependence has been proposed elsewhere \cite{Chowdhury2015} but no quantitative calculation has been conducted until now. To understand this, one recalls that in conventional Kondo models, the Kondo scale depends strongly on the hybridization function around the Fermi level. Because the resonance width depends on the hybridization spectrum at the Fermi level $\Delta_r \sim -\frac{1}{\pi} \mathrm{Im} \Delta(\omega = 0)$, the Kondo physics for symmetric hybridization spectrum can be observed if $-U - \Delta_r < \epsilon_d < \Delta_r$ \cite{Krishna-murthy1980b,Hewson1997}. For the pseudogap model, $\Delta_r = 0$, the necessary condition to observe the Kondo physics is $-U < \epsilon_d < 0$. However, because there is no phase transition at half-filling for $r=1$ pseudogap model, the line $\epsilon_d = -\frac{U}{2}$ is the upper limit for all phase boundary curves in this region. Besides, the local moment screening effect is enhanced by the hybridization strength $v_0$ and the valence mixing controlled by $\epsilon_d$. Therefore, at large $v_0$, a small amount of valence mixing is enough to screen the impurity local moment, which means $\epsilon_d^c$ is closer to $-\frac{U}{2}$ as $v_0$ increases.

Interestingly, we find that in Fig.~\ref{fig:phase_diagram}(c) the phase boundary plotted in the $(U, \epsilon_d)$ plane exhibits linear behavior
\begin{equation}\label{eq:slope}
U^c = -\alpha_t \epsilon_d^c,
\end{equation}
where $\alpha_t$ is defined as the slope of the phase boundary, which depends strongly on $v_0$. We will study several aspects of this linear feature. First, we examine the dependence of the $(U^c, \epsilon_d^c)$ line on $v_0$. Second, we compare it with that of low-energy pseudogap impurity models which are often used in literature, i.e. models where the DOS of the host is $\nu(\omega) \sim |\omega|^r$ \cite{Withoff1990,Gonzalez-Buxton1998}.

\begin{figure}
 \includegraphics[width=\columnwidth]{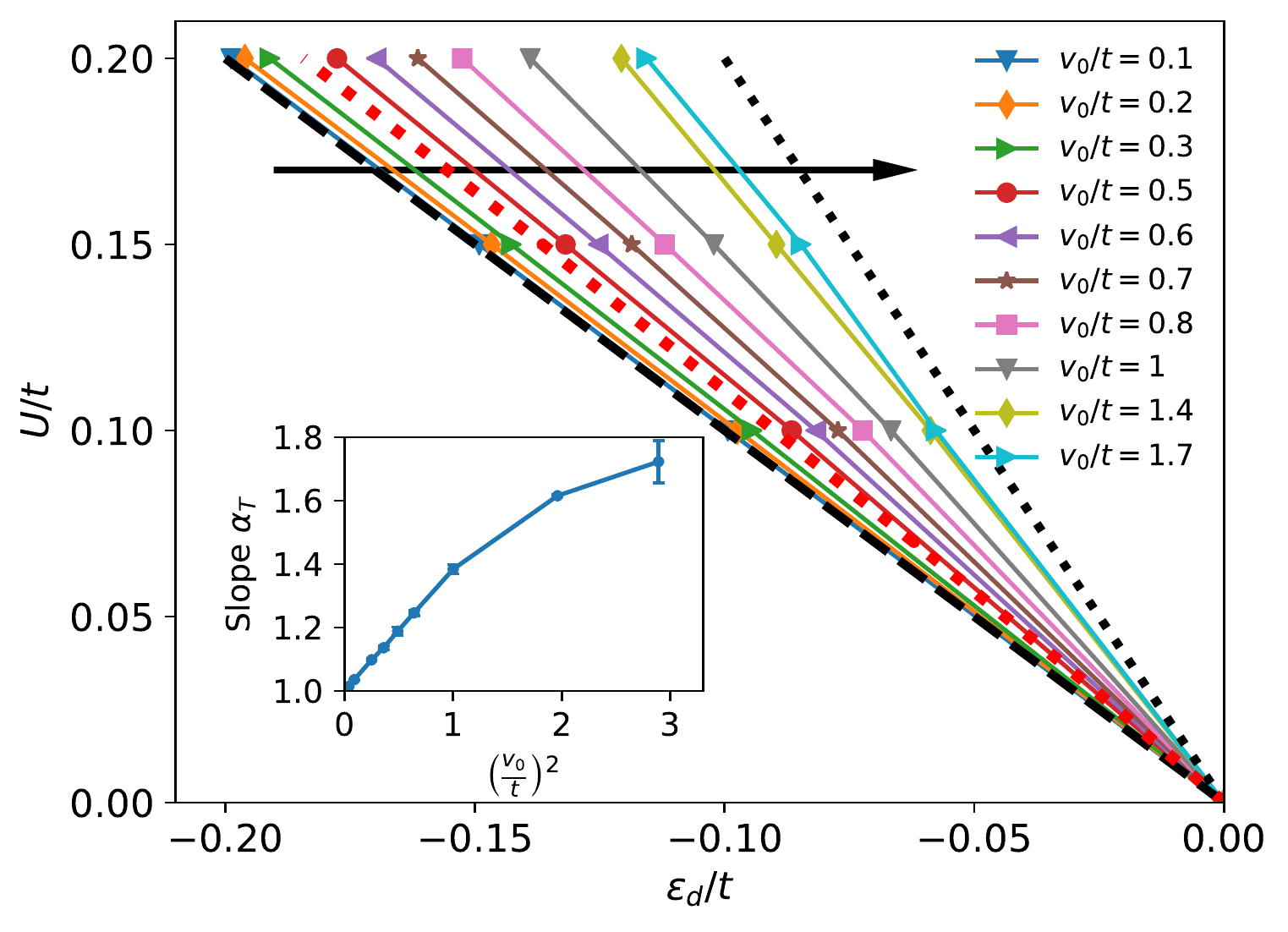}
\caption{\label{fig:slopes} The phase boundaries in the $(U, \epsilon_d)$ plane plotted for various $v_0$ values. Lines of increasing $v_0/t$ appear along the direction of the black solid arrow. The black dashed line is the lower limit $\epsilon_d = -U$, the black dotted line is the upper limit $\epsilon_d = -U/2$. The red dotted line is the phase boundary for the conventional pseudogap Anderson model [see Eq.~\eqref{eq:pixley_hyb}] which has the same hybridization spectrum slope as the honeycomb lattice model at $v_0 = 0.5t$. Inset: the slope $\alpha_t$ of the $U^c$ vs. $\epsilon_d^c$ line depending on $(v_0/t)^2$. The error bars of $\alpha_t$ in the inset are comparable to the symbol size except for the point of largest $v_0$.}
\end{figure}

In Fig.~\ref{fig:phase_diagram}(a), the three phase boundaries are equally distanced as $U$ increases but the differences between the three $\epsilon_d^c$ decreases as $v_0$ increases, implying the dependence of $\alpha_t$ on $v_0$. Based on the data for Fig.~\ref{fig:phase_diagram}(a) and the fact that the phase boundary line goes towards the origin of the $(U, \epsilon_d)$ plane, we plot in Figure~\ref{fig:slopes} the phase boundaries for various $v_0$ values. The phase boundaries approach $\epsilon_d = -U$ (the black dashed line) illustrating the lower limit described above. They are all below the upper limit $\epsilon_d = -\frac{U}{2}$ (the black dotted line). However it requires larger $v_0$ to observe the approach to the upper limit.

Because the Kondo effect is a low-energy physical phenomenon, we expect that even in the honeycomb lattice (graphene), the physics is only affected by changes in proximity to the Fermi level. In our case, the only change in low energy ($\omega \sim 0$) is the slope of the hybridization spectrum. Around the Fermi level, the tight-binding DOS of graphene is linear $\nu(\omega) = \frac{A_c}{\pi v_F^2}|\omega|$ (where $A_c = \frac{3\sqrt{3}}{2}$ is the area of the unit cell and $v_F=\frac{3t}{2}$ is the Fermi velocity) \cite{Neto2009}. Employing Eq.~\eqref{eq:t_hyb}, the slope $\gamma_t$ of the hybridization spectrum (a dimensionless parameter) depends only on $v_0^2$
\begin{equation}\label{eq:gamma_t}
 \gamma_t = \dfrac{v_0^2}{2} \dfrac{A_c}{\pi v_F^2} = \dfrac{v_0^2 \sqrt{3}}{3 \pi t^2}.
\end{equation}

We verify our statement by comparing our results for honeycomb lattice with those from standard pseudogap problem, in which the hybridization spectrum is set to have the same slope
\begin{equation}\label{eq:pixley_hyb}
 -\dfrac{1}{\pi} \mathrm{Im} \Delta(\omega) = \gamma_t |\omega| \theta(|\omega| - D).
\end{equation}
Here the step function $\theta(x)$ limits the spectra to be nonzero only in the range $(-D, D)$, defining the bandwidth $2D$. For simplicity, in this test we set $D=t$. We test with $v_0 = 0.5t, 0.7t$ and $1.4t$. In Fig.~\ref{fig:slopes} the phase boundary of the conventional model corresponding to $v_0 = 0.5t$ (the red dotted line) is plotted for comparison. This conventional pseudogap phase boundary stays rather closely to the $v_0=0.5t$ phase boundary of the honeycomb lattice model, the difference between the two phase boundary slopes is $4\%$. For other cases corresponding to $v_0 = 0.7t$ and $1.4t$ (not shown), the differences are $6\%$ and $5\%$, respectively. In addition to the error in CT-HYB measurements, the difference of the phase boundaries between the honeycomb lattice model and the simplifed pseudogap model might be due to the higher energy effect that may involve, resulting in the contribution of nonlinear higher order terms in the tight-binding honeycomb lattice DOS. Nevertheless, the difference is small, thus the similarity between the two models provides evidence that, because the Kondo effect is a low-energy effect, the physics near the Fermi level plays a crucial role. In this case, it is represented by the slope of the hybridization spectra, which depends mostly on $\frac{v_0}{t}$.

In the inset of Fig.~\ref{fig:slopes}, the dependence of the slope $\alpha_t$ of the phase boundary on $\left(\frac{v_0}{t}\right)^2$ is plotted. Interestingly, this slope depends linearly on $v_0^2$ for small $v_0$ ($v_0 < t$) and approaches the lower limit value $1$ as $v_0\to 0$ (corresponding to $\epsilon_d^c = -U$). For $v_0 > t$, $\alpha_t$ slowly approach the upper limit value $2$ (the upper limit at half-filling $\epsilon_d = -\frac{U}{2}$) as $v_0$ increases. Thus $v_0$ is the key factor that controls the slope $\alpha_t$ of the $(U^c,\epsilon_d^c)$ phase boundary. At different $v_0$ scales, $\alpha_t$ behaves differently.

\section{Analyze the phase diagram \label{sec:analyze}}

We will discuss in details the behavior of each type of phase boundary in the phase diagram presented in Fig.~\ref{fig:phase_diagram}. In these phase boundries, there are two interesting features for further analysis: (1) the linear relation between $U^c$ and $\epsilon_d^c$ for fixed $v_0$, and (2) the dependence of the slope $\alpha_t = -U^c/\epsilon_d^c$ as well as $\epsilon_d^c$ itself on $v_0$. The impurity self-energy is used intensively to understand these features.

\begin{figure}
 \includegraphics[width=\columnwidth]{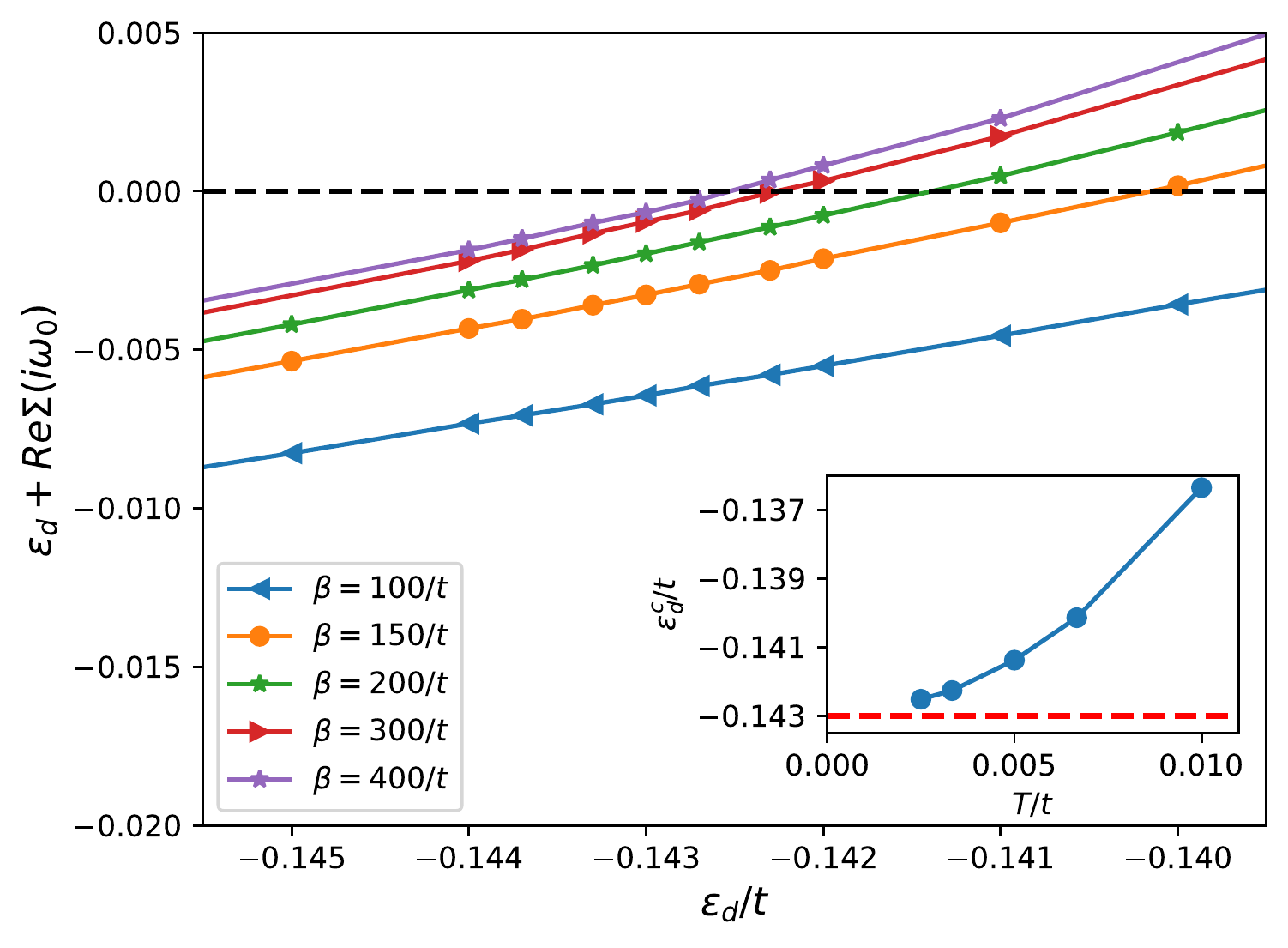}
\caption{\label{fig:self_energy} The plot of $y = \epsilon_d + \mathrm{Re}\Sigma(i\omega_0)$ vs. $\epsilon_d$ (where $\omega_0 = \pi/\beta$ is the first Matsubara frequency) for $U=0.15t$ and $v_0=0.3t$. The solution $\epsilon_d^c(T)$ is determined from the crossing point of the curves with the horizontal black dashed line ($y=0$). Inset: Temperature dependence of $\epsilon_d^c(T)$. The red horizontal dashed line marks the critical value $\epsilon_d^c$ obtained from the Binder analysis.}
\end{figure}

\subsection{Impurity self-energy}

In the Kondo model, the self-energy around the Fermi level is the first signature for the correlation effects. We state that the Kondo physics, if occurs, must be accompanied with the Kondo peak of the impurity DOS around the Fermi level \cite{Hewson1997}. Otherwise, the Kondo peak disappears and there is only a more broadened peak as a result of the hybridization between the impurity energy level and the conduction band. Therefore, the onset of the quasiparticle excitations representing the Kondo peak is the signature for the QCP. The equation for the quasiparticle excitation energy reads
\begin{equation}
 \omega - \epsilon_d - \mathrm{Re}\Sigma_R(\omega) - \mathrm{Re}\Delta(\omega) = 0.
\end{equation}
where $\Sigma_R(\omega)$ is the retarded self-energy of the impurity. Around the Fermi level $\omega \to 0$, $\mathrm{Re}\Delta(\omega) \to 0$ because the pseudogap hybridization spectrum is symmetric around the Fermi level. We thus obtain the criterion for the quantum phase transition
\begin{equation}\label{eq:crit_sigma}
 \epsilon_d^c + \mathrm{Re}\Sigma_R(\omega\to0) = 0.
\end{equation}
In practice, $\mathrm{Re}\Sigma_R(\omega\to0)$ can be approximated by the Matsubara self-energy at the lowest frequency $i\omega_0 = i \frac{\pi}{\beta}$. In Figure~\ref{fig:self_energy} the dependence of $\epsilon_d + \mathrm{Re}\Sigma(i\omega_0)$ on $\epsilon_d$ at various temperatures is plotted for the case of $U = 0.15t$ and $v_0 = 0.3t$ to demonstrate the criterion in Eq.~\eqref{eq:crit_sigma}. The crossing point $\epsilon_d^c(T)$ of each curve with the $y=0$ black dashed line  is an estimate of the critical $\epsilon_d^c$ at the corresponding temperature. Decreasing the temperature, the inset of Fig.~\ref{fig:self_energy} shows that $\epsilon_d^c(T)$ approaches the critical value $\epsilon_d^c$ obtained from the Binder analysis, justifying the critical condition in Eq.~\eqref{eq:crit_sigma}. Therefore $\mathrm{Re}\Sigma(i\omega_0)$ can be used to understand further the physics of the phase boundary.

\begin{figure}
 \includegraphics[width=\columnwidth]{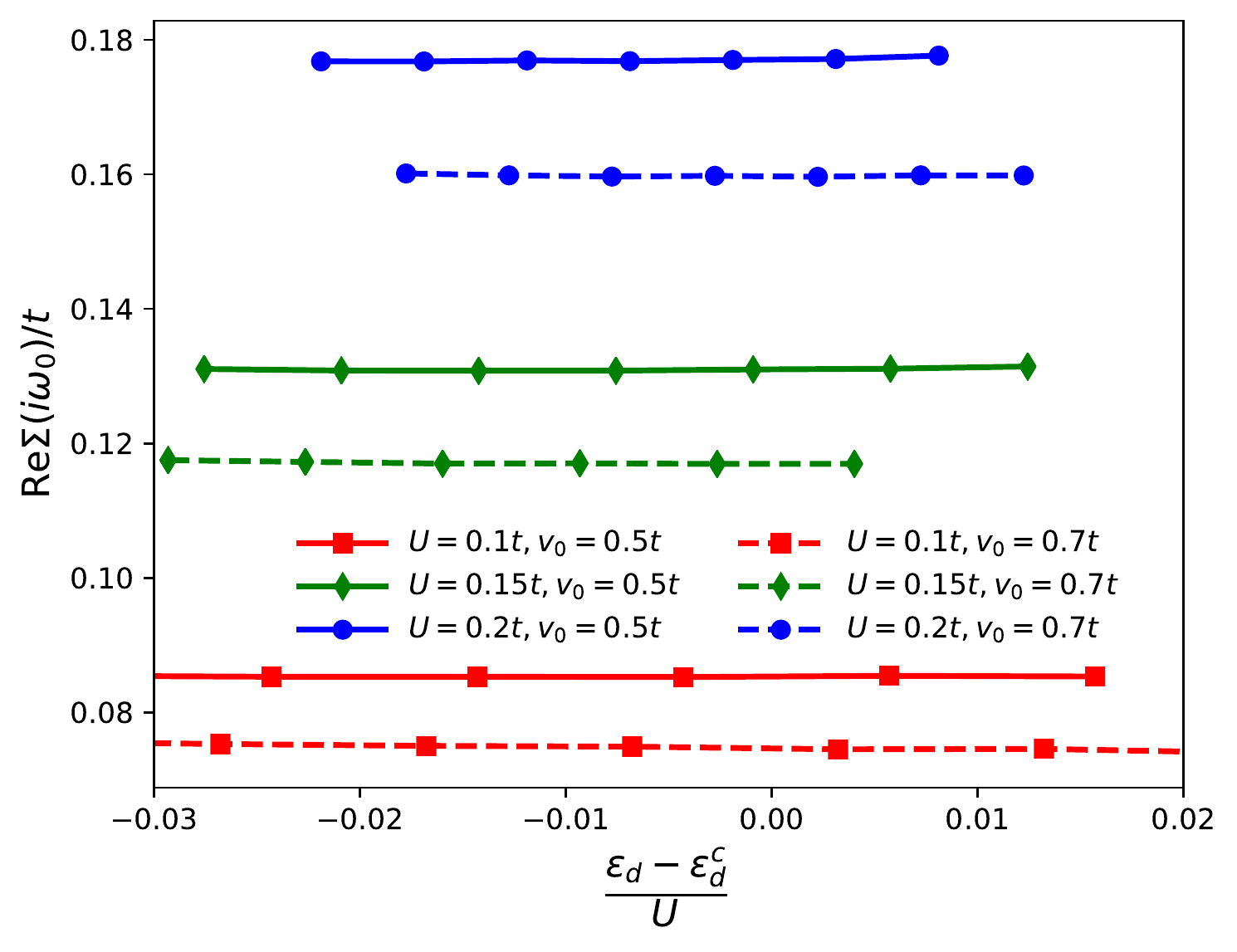}
\caption{\label{fig:se_stable} Real part of the self energy $\mathrm{Re}\Sigma(i\omega_0)$ depending on $\frac{\epsilon_d - \epsilon_d^c}{U}$ plotted for several sets of $(U, v_0)$. Solid lines (dashed lines) correspond to $v_0=0.5t$ ($v_0=0.7t$). Lines of the same color and symbol are for the same $U$.}
\end{figure}

We first examine $\mathrm{Re}\Sigma(i\omega_0)$ at the lowest temperature used in our simulation ($T = \frac{t}{400}$) for different values of $\epsilon_d$, $U$ and $v_0$. Figure~\ref{fig:se_stable} exhibits the dependence of $\mathrm{Re}\Sigma(i\omega_0)$ on $\epsilon_d$ for different sets of $(U, v_0)$ around the critical values $\epsilon_d^c$. It shows that when $\epsilon_d$ is varied around the critical region (in the range of a few percent of $U$ around $\epsilon_d^c$), $\mathrm{Re}\Sigma(i\omega_0)$ is nearly constant (changed by about $0.5\%$). Thus, for each case of $(U, v_0)$ one can estimate $\mathrm{Re}\Sigma(i\omega_0)$ at criticality without the need for a highly accurate $\epsilon_d^c$. In Fig.~\ref{fig:se_stable}, $\mathrm{Re}\Sigma(i\omega_0)$ behaves similarly to $\epsilon_d^c$, it changes linearly with respect to $U$ when keeping $v_0$ fixed. When we increase $v_0$ from $0.5t$ to $0.7t$, this linear relation is changed in the same way as that of $U^c$ vs. $\epsilon_d^c$ [see Fig.~\ref{fig:phase_diagram}(a)]. Therefore, despite at nonzero temperature, $\mathrm{Re}\Sigma(i\omega_0)$ reflects reasonably the value of $\epsilon_d^c$, justifying Eq.~\eqref{eq:crit_sigma}.

\subsection{Linear relation between $U^c$ and $\epsilon_d^c$}

\begin{figure}
 \includegraphics[width=\columnwidth]{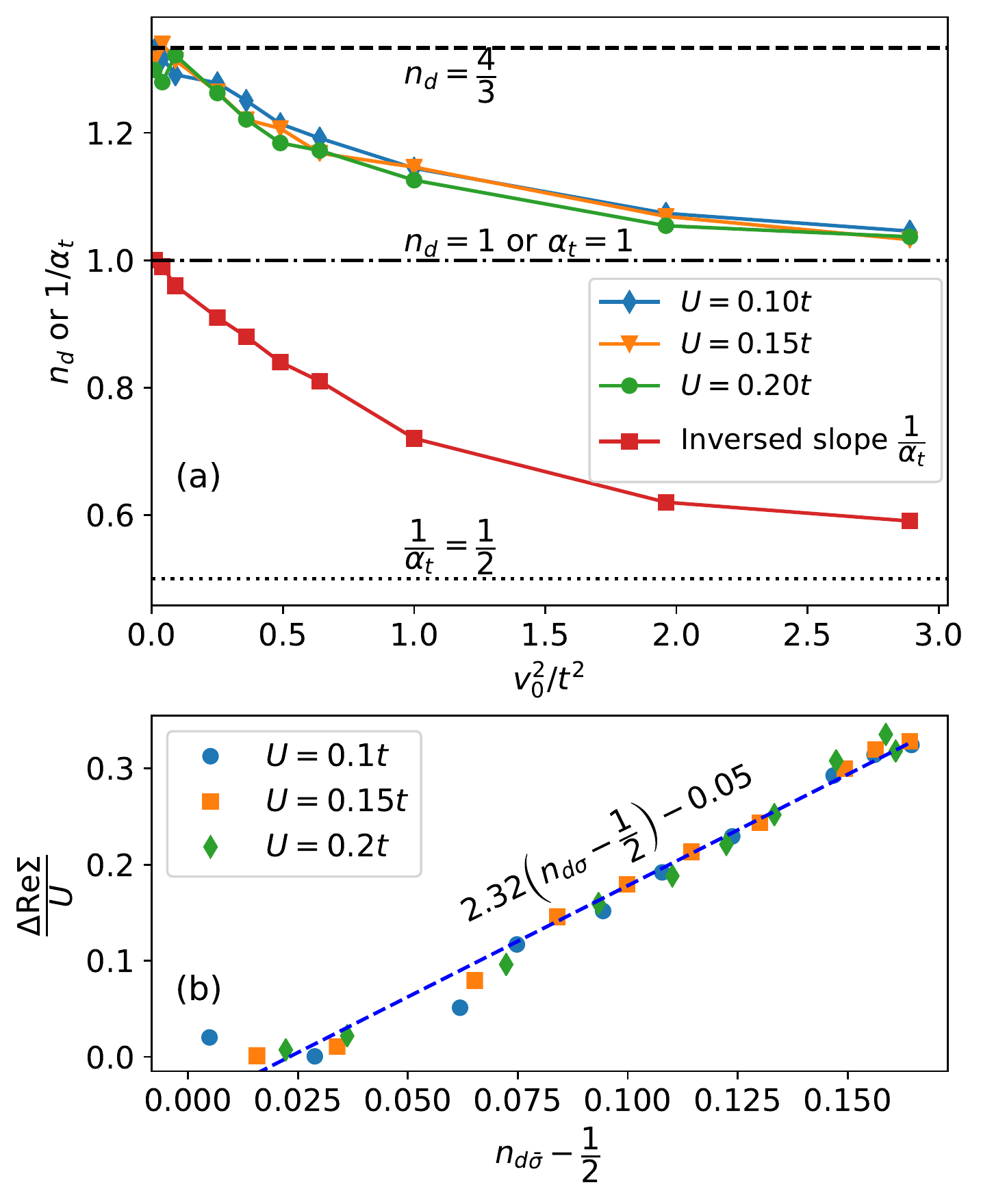}
\caption{\label{fig:nd_sigma} (a) Impurity occupancy vs. $v_0^2$ for $U=0.10t, 0.15t$ and $0.20t$. The inversed slope $\frac{1}{\alpha}$ is also plotted for comparison. The black dashed line marks the limit of $n_d$ as $v_0 \to 0$. The black dashed dotted line marks the limit of $n_d$ as $v_0 \to \infty$ and the limit of $\alpha$ as $v_0 \to 0$. The black dotted line marks the limit of $\alpha_t$ as $v_0 \to \infty$. (b) The dependence of $\frac{\Delta\mathrm{Re}\Sigma}{U}$ on $n_{d\bar{\sigma}} - \frac{1}{2}$ for three $U$ values. The blue dashed line is the least-square fitting line with the slope $2.33(7)$ and the intercept $-0.056(8)$.}
\end{figure}

From the mean field theory for the Anderson impurity model, the self-energy is approximated by the Hartree term $\Sigma^H_\sigma = U n_{d\bar{\sigma}}$ where $n_{d\sigma} = \langle \hat{n}_{d\sigma}\rangle$ is the impurity occupancy per spin. The mean-field theory still has an impact even in the solution of the flatband symmetric Anderson model where the real part of the self-energy at zero frequency is exactly equal to the Hartree term $\mathrm{Re}\Sigma_\sigma(\omega \to 0) = U n_{d\bar{\sigma}}$ \cite{Horvatic1980,Bulla1998}. Extending to strongly correlated lattice systems, previous works \cite{Dang2014a,Dang2014b} suggest that the correlated electron occupancy characterize the correlation strength of a system based on how localized the correlated electrons are. Unless a phase transition occurs, properties such as spectral functions may only experience minor changes when the interaction $U$ is increased as long as $n_{d\sigma}$ remains unchanged. Thus it encourages us to examine the impurity occupancy $n_d = n_{d\uparrow} + n_{d\downarrow}$ at criticality, which is estimated by interpolating the occupancy for various $\epsilon_d$ around the critical point. From Eqs.~\eqref{eq:slope} and \eqref{eq:crit_sigma}, the relation $\mathrm{Re}\Sigma(\omega\to 0) = \frac{1}{\alpha_t} U$ means that the inversed slope $1/\alpha_t$ might be closely related to $n_d$. Figure~\ref{fig:nd_sigma}(a) shows $n_d$ in comparison with $1/\alpha_t$ for a wide range of $v_0$ extracted from various $(U^c, \epsilon_d^c)$ phase boundaries. We note that in the atomic limit, $\frac{dn_d}{d\epsilon_d}$ becomes large, thus the value of $n_d$ at $v_0 \to 0$ contains large error bar, causing the noise of $n_d$ at small $v_0$ in Fig.~\ref{fig:nd_sigma}(a).

Fig.~\ref{fig:nd_sigma}(a) gives us useful information. First, the two limits of $v_0$ for the plot can be calculated exactly. In the atomic limit $v_0\to 0$, $n_d = \frac{4}{3}$ and $\alpha_t = 1$; whereas at $v_0\to \infty$, the impurity reaches half-filling, $n_d = 1$ and $\alpha_t = 2$. In our simulations, both $n_d$ and $1/\alpha_t$ approach these limits, although it may require simulations of larger $v_0$ to observe more clearly the asymtotic behavior at $v_0\to\infty$. Second, $1/\alpha_t$ varies in a similar way as $n_d$, suggesting that the real part of the self-energy behave similarly to the Hartree term $|\epsilon_d^c| = \mathrm{Re}\Sigma_\sigma(\omega\to 0) \sim Un_{d\bar{\sigma}}$. It motivates us to investigate the difference between $\mathrm{Re}\Sigma(\omega\to 0)$ and the Hartree self-energy
\begin{equation}
 \Delta\mathrm{Re}\Sigma = \mathrm{Re}\Sigma_\sigma(\omega\to 0) - U n_{d\bar{\sigma}}.
\end{equation}
We argue that, as $n_d$ behaves similarly to $1/\alpha_t$ and is weakly dependent on $U$ and the self-energy reaches Hartree value as the impurity reaches half-filling, $\Delta\mathrm{Re}\Sigma$ can be an explicit function of $U$ and $n_{d\bar{\sigma}} - 1/2$, i.e. $\Delta\mathrm{Re}\Sigma = f(U, n_{d\bar{\sigma}} - 1/2)$. It suggests that at critical points one rescale $\Delta\mathrm{Re}\Sigma$ by $U$ and plot with respect to $n_{d\bar{\sigma}} - 1/2$ so that both $x$ and $y$ axes are dimensionless. Interestingly, Figure~\ref{fig:nd_sigma}(b) shows that, in such a plot, points of three different $U$'s seem to collapse within numerical error bars. Therefore, we may express $\mathrm{Re}\Sigma(\omega\to 0)$ at criticality by an empirical formula
\begin{equation}\label{eq:nd_sigma}
 \mathrm{Re}\Sigma_\sigma(\omega\to 0) = U n_{d\bar{\sigma}} + c U (n_{d\bar{\sigma}} - \dfrac{1}{2}),
\end{equation}
where $c = 2.32(7)$ is the slope of the fitting line in Fig.~\ref{fig:nd_sigma}(b). In the limit of $v_0 \to 0$, $n_{d\sigma} \to \frac{2}{3}$ and $\mathrm{Re}\Sigma(\omega\to 0) \to U$, then $c = 2$. The difference between numerical and asymptotic results of $c$ might be understood if simulations at small $v_0$ could be conducted, which is however unavailable in this work. As a result, given Eq.~\eqref{eq:nd_sigma} and using $c = 2$, $\frac{1}{\alpha_t}$ is expressed in terms of $n_d = 2 n_{d\sigma}$ as
\begin{equation}
 \dfrac{1}{\alpha_t} = \dfrac{1+c}{2}n_d - \dfrac{c}{2} = \dfrac{3}{2}n_d - 1.
\end{equation}

We come to an interpretation that at criticality, $n_d$ tends to determine the screening of the impurity interaction, in similarity to lattice correlated systems \cite{Dang2014a,Dang2014b}. When $U$ is increased, $n_d$ tends to be fixed in order to keep the physics at criticality remains unchanged, explaining the weak dependence of $n_d$ on $U$, as seen in Fig.~\ref{fig:nd_sigma}(a). As $\mathrm{Re}\Sigma(\omega\to0)$ can be expressed purely by $U$ and $n_d$, it explains the linearity of the phase boundaries in the  $(U, \epsilon_d)$ plane as plotted in Fig.~\ref{fig:slopes}.

\subsection{The role of $v_0$}

Different from the $U^c$ vs. $\epsilon_d^c$ linear relation discussed in the previous subsection, the dependence of $\epsilon_d^c$ and $\alpha_t$ on $v_0$ is nonlinear. From Eqs.~\eqref{eq:delta} and \eqref{eq:t_hyb}, $v_0$ enters the hybridization formula as $v_0^2$. Thus one can think of expanding $\mathrm{Re}\Sigma(\omega\to 0)$ with respect to $v_0^2$ ($1/v_0^2$) at small (large) $v_0$. Indeed, one could employ strong- or weak-coupling perturbation theories \cite{Dai2005,Yosida1975,Horvatic1980} (with $\epsilon_d$ adjusted to equals $\mathrm{Re}\Sigma(\omega\to0)$) to estimate $\mathrm{Re}\Sigma(\omega\to0)$ so as to explain the numerical results in the limit of small and large $v_0$ presented in Sec.~\ref{sec:t_pos}. It is however beyond the scope of this work. In this subsection, we attempt to obtain an empirical formula for the self-energy, which serves as the suggestion for future perturbative calculations of this model.

From Eqs.~\eqref{eq:slope} and \eqref{eq:crit_sigma}, the self-energy can be estimated based on the slope of the $(U^c,\epsilon_d^c)$ phase boundary
\begin{equation}\label{eq:resigma}
 \mathrm{Re}\Sigma(\omega \to 0) = \dfrac{U^c}{\alpha_t}.
\end{equation}
We then expand $\alpha_t$ with respect to $v_0^2$ ($1/v_0^2$) at small (large) $v_0$. Based the inset of Fig.~\ref{fig:slopes}, we expect that, at small $v_0$, 
\begin{equation}
\alpha_t \approx 1 + a v_0^2,
\end{equation}
while at large $v_0$
\begin{equation}
\alpha_t \approx 2 - \dfrac{4b}{v_0^2},
\end{equation}
where $a$ and $b$ are fitting parameters. The constant terms $1$ and $2$ are the values at the limits $v_0 \to 0$ and $v_0 \to \infty$, respectively. After fitting, the results are $a = 0.383(2)/t^2$ and $b = 0.19(1)t^2$. Replace $a$ and $b$ into Eq.~\eqref{eq:resigma} and use the approximation $\dfrac{1}{1 \pm x} \approx 1 \mp x$ for $x \ll 1$, we obtain the empirical formulas for the real part of the self-energy at zero frequency at critical points
\begin{equation}\label{eq:emp_sigma}
\begin{aligned}
 \mathrm{Re}\Sigma(\omega \to 0) &= U^c (1 - a v_0^2) & \text{if } v_0 \to 0, \\
 \mathrm{Re}\Sigma(\omega \to 0) &= U^c \left(\dfrac{1}{2} + \dfrac{b}{v_0^2} \right)  & \text{if } v_0 \to \infty,
\end{aligned}
\end{equation}

We note that the parameters $a$ and $b$ can be fitted directly using the self-energy data $\mathrm{Re}\Sigma(i\omega_0)$ or the critical values $\epsilon_d^c$ at different $v_0$. However, it turns out that fitting using $\mathrm{Re}\Sigma(i\omega_0)$ or $\epsilon_d^c$ requires rather small $\frac{v_0}{U}$ (large $\frac{U}{v_0}$) to estimate $a$ ($b$), which exceeds our simulation capability. With the data used to construct Fig.~\ref{fig:phase_diagram}, fitting the self-energy with $v_0 < 0.7t$ gives $a = 0.55(1)/t^2, 0.48(1)/t^2$ and $0.44(1)/t^2$ for $U=0.1t, 0.15t$ and $0.2t$; with $v_0 > 0.8t$, it gives $b = 0.11(2)t^2, 0.15(3)t^2$ and $0.17(3)t^2$. These fittings are thus unstable, showing clear dependence of $a$ and $b$ on $U$. Obtaining $a$ and $b$ by fitting the slope $\alpha_t$ is more advantageous because it exploits the linear relation between $U^c$ and $\epsilon_d^c$, thus neglecting the dependence on $U^c$. The data for $v_0$ in the range from $0.1$ to $1.7$ together with the limit $v_0 \to 0$  and $v_0 \to \infty$ is enough for fitting $\alpha_t$. Hence, the results $a = 0.383(2)/t^2$ and $b = 0.19(1)t^2$ extracted from the slope $\alpha_t$ are more reliable.

Therefore, we present in Eq.~\eqref{eq:emp_sigma} empirical formulas for $\mathrm{Re}\Sigma(\omega\to 0)$ corresponding to $v_0\to 0$ and $v_0 \to \infty$. These formulas can be the initial guideline for future perturbative calculations, which might help to understand further the dependence of the phase boundaries on $v_0$.

\section{At hollow and bridge sites\label{sec:bh_pos}}

So far, we focus mostly on the situation where the impurity is placed on top of an atom on the honeycomb lattice. This is a special case in which the hybridization spectrum is symmetric around the Fermi level and there exists particle-hole symmetry at $\epsilon_d = -\frac{U}{2}$. In general, the hybridization spectrum depends on the impurity position in the lattice. In Appendix~\ref{app:ph_symmetry}, we show that when the impurity departs from this $t$ site, it hybridizes with atoms of both sublattices, the hybridization spectrum is no longer symetric around the Fermi level. When the hybridization spectrum changes, it obviously affects the physics. In this section, we discuss briefly the effects when the impurity is placed at the hollow $h$ and at the bridge $b$ sites [see Fig.~\ref{fig:imp_graphene}].

\begin{figure}
 \includegraphics[width=\columnwidth]{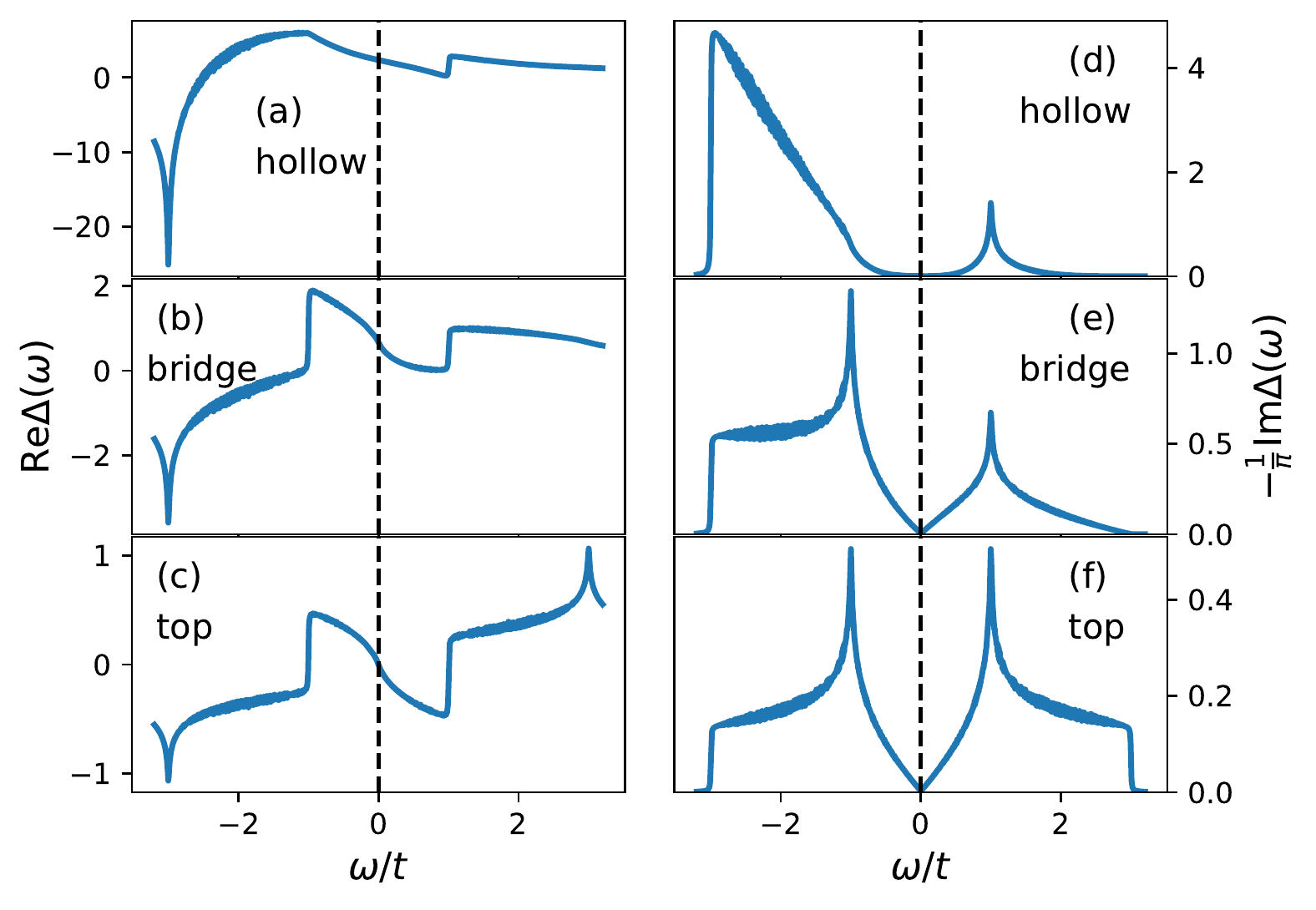}
\caption{\label{fig:hyb_dos} The hybridization function in real frequency for the same $v_0$ (in this plot $v_0 = t$) for three impurity positions (at hollow $h$, at the bridge $b$ and on top of an atom $t$). The left panel is the real part $\mathrm{Re}\Delta(\omega)$. The right panel is the hybridization spectrum $-\frac{1}{\pi}\mathrm{Im}\Delta(\omega)$.}
\end{figure}

Figure~\ref{fig:hyb_dos} shows the real part and the spectrum $-\frac{1}{\pi}\mathrm{Im}\Delta(\omega)$ of the hybridization function when the impurity is placed at $t$, $h$ and $b$ positions for the same $v_0$. Because the hybridization function is directly proportional to $v_0^2$, for simplicity, we fix $v_0/t = 1$ in the figure. For different $v_0$, one simply multiplies the hybridization function with $v_0^2$ to obtain a new one. Fig.~\ref{fig:hyb_dos} shows that $\mathrm{Re}\Delta(\omega=0)$ increases as the impurity changes from $t \to b \to h$. Due to contributions from both sublattices (see Appendix~\ref{app:ph_symmetry}), the imaginary parts obviously show the asymmetry of the hybridization spectra when the impurity is not at the $t$ site. We have found that while the hybridization spectra for the $t$ and $b$ sites remain linear around the Fermi level, it behaves as $|\omega|^3$ for the $h$ site case. Expansion around the Fermi level shows the forms of $\Delta(\omega)$ around $\omega = 0$, of which the results are summarized in Table~\ref{table:hyb_spectra}.

Nonzero $\mathrm{Re} \Delta(\omega=0)$ values of impurity at the $b$ and $h$ sites result in the shifting of the critical impurity level $\epsilon_d^c$. By defining an effective energy level $\epsilon_{eff} = \epsilon_d + \mathrm{Re}\Delta(\omega=0)$, the critical value $\epsilon_{eff}^c$ is in the same energy range as $\epsilon_d^c$ for impurity at the $t$ site $(-U < \epsilon_{eff}^c < -\frac{U}{2})$, allowing to determine $\epsilon_d^c$ in these cases more easily. The difference in the real part of $\Delta(\omega)$, of which the main effect is to introduce a shifting in $\epsilon_d^c$, is thus not as important as the change in the imaginary part.

When placing the impurity at the $b$ site, the hybridization function exhibits the asymmetric spectrum, starting from the the second order of the frequency
\begin{equation}\label{eq:hyb_b}
 -\dfrac{1}{\pi} \mathrm{Im}\Delta(\omega) =
 \dfrac{v_0^2 A_c}{\pi v_F^2}\left(|\omega| - \dfrac{\omega^2}{2 v_F} sgn(\omega) + o(\omega^3)\right).
\end{equation}
Examining the $(U, \epsilon_d)$ phase diagram at $v_0 = 0.5t$ for the tight-binding honeycomb lattice Eq.~\eqref{eq:delta} and for the simplified model in Eq.~\eqref{eq:hyb_b}, the phase boundaries (not shown) still exhibit linear relation. The difference of the boundary slope between the lattice model and the simplified model is about $2\%$, which we argue that it is due to the differences in the hybridization function at frequencies away from $\omega = 0$. Apart from that, we do not observe significant changes when changing the location of the impurity from the $t$ site to the $b$ site. Thus, although the hybridization spectrum is asymmetric, as long as the spectrum still exhibits linear behavior around the Fermi level, there is no significant change in physics.

\begin{table}[t]
\begin{ruledtabular}
\begin{tabular}{ c  c  c  c  }
          & $t$    & $b$   & $h$     \\
\hline
$\mathrm{Re}\Delta(\omega = 0)$ & $0$    & $\sim 0.66 v_0^2$ & $\sim 2.35 v_0^2$ \\
$-\dfrac{1}{\pi} \mathrm{Im}\Delta(\omega)$ & $\dfrac{v_0^2 A_c}{2\pi} \dfrac{|\omega|}{v_F^2}$ & $\dfrac{v_0^2 A_c}{\pi} \dfrac{|\omega|}{v_F^2}$ & $\dfrac{9 v_0^2 A_c}{4\pi} \dfrac{|\omega|^3}{v_F^4}$ 
\end{tabular}
\end{ruledtabular}
\caption{ \label{table:hyb_spectra} The real part of the hybridization function at $\omega = 0$ (calculated numerically) and the hybridization spectrum $-\frac{1}{\pi}\mathrm{Im}\Delta(\omega)$ at the lowest order of $\omega$ for impurity located at $t$, $b$ and $h$ positions. $A_c = \frac{3\sqrt{3}}{2}$ is the area of the unit cell, $v_F = \frac{3t}{2}$ is the Fermi velocity.}
\end{table}

Impurity at the $h$ site is a different scenario. In this case, the impurity strongly hybridizes with conduction electrons of six nearest neighbor sites. We find that $|\omega|$ and $\omega^2$ terms in the hybridization spectrum vanish, leaving the $|\omega|^3$ term as the lowest order term in the spectrum (see Table~\ref{table:hyb_spectra}). The impurity problem is transformed into an $r=3$ pseudogap Kondo model. Due to very small hybridization spectral weight around the Fermi level, it turns out to be not easy to observe the Kondo screening phase or determine the critical point with high accuracy in this case. Indeed, our numerical calculations contain large error bars, which prohibit us from reliable extrapolation for $\epsilon_d^c$.

Therefore, the physics of the pseudogap Kondo model of an impurity on the honeycomb lattice depends on the location of the impurity on the lattice. If it is placed at $b$ and $t$ sites, linear behavior of the host DOS is retained in the hybridization spectrum, allowing for the observation of the Kondo screening phase. However, around the $h$ site, the hybridization spectrum near the Fermi level is nearly depleted and becomes a different type of pseudogap model. At this impurity position, the conditions turn out to be more extreme in order to observe the Kondo physics.

\section{Conclusions \label{sec:conclusions}}

In this paper, we have studied the pseudogap Anderson model for an $s$-wave impurity placed on a graphene-like honeycomb lattice. We used the tight binding model to describe conduction electrons of the host material, and employed the hybridization-expansion continuous-time quantum Monte Carlo method (CT-HYB) to fully treat the correlation effect. We focused on the case the impurity is placed on top of an atom in the lattice (the $t$ position) and constructed the phase diagram with three parameters $U, v_0$ and $\epsilon_d$ for the quantum phase transition from the free local moment phase to the Kondo screening phase. We analyzed the impacts of the hybridization and the correlation strength on the phase diagram and provided insightful discussion for understanding of the phase diagram based on the self-energy of the impurity. We also discussed qualitatively the physics and the possibility to observe the Kondo screening phase when the impurity position is moved away from $t$ position.

The work presents calculations from high-temperature viewpoint and contains several interesting results. First, we have found that the full phase diagram exhibits interesting relations between each pair of $(U^c, \epsilon_d^c)$ and $(v_0, \epsilon_d^c)$. The $(U^c, \epsilon_d^c)$ phase boundary is linear for the range of $v_0$ under investigation; while in the $(v_0, \epsilon_d^c)$ plane, $\epsilon_d^c$ exhibits $v_0^2$ (or $\frac{1}{v_0^2}$) dependence in the limit of small (large) $v_0$. We have discussed the results by relating the critical value $\epsilon_d^c$ to the impurity self-energy $\Sigma(i\omega_n)$ and obtained an empirical expression for the self-energy in order to gain more insights into the physics of the phase diagram. Lastly, we have found that changing the impurity from position on top of an atom to the bridge ($b$) or hollow ($h$) site might affect the physics. Specically, while the system remains in the $r=1$ pseudogap Anderson model if the impurity is at the $t$ or $b$ sites, it becomes an $r=3$ pseudogap model if the impurity is placed at the $h$ site.

In overall, the realization of the $r=1$ pseudogap model in graphene is still a challenging task for both experimental and theoretical sides. Indeed, the transition from the free local moment to Kondo screening phase depends on the position of the impurity on the honeycomb lattice. The prominent case to observe the Kondo physics is when the impurity is located at the $t$ site, where the hybridization spectrum is symmetric and directly proportional to the DOS of the honeycomb lattice. If it is located at the $h$ site, it becomes a different pseudogap model and is not easy to observe the Kondo screening effect. Although the hybridization at the hollow site for a more realistic $d$-wave impurity might be slightly different \cite{Wehling2010a,Wehling2010b}, the small slope of the hybridization spectrum around the Fermi level remains and is an obstacle for any measurement of the Kondo screening phase.

Our work is restricted to the CT-HYB impurity solver, which is a high-temperature method, with an assumption of a single $s$-wave impurity for the sake of simplicity. Therefore, certain limitations in the work are inevitable. The major one is that, due to the CT-HYB impurity solver, we cannot access low temperature region in order to enter the quantum critical regime and measure critical behaviors more accurately. There is also a lack of analytic calculations for asymtotic limits, such as at $v_0\to 0$ and $v_0 \to \infty$, in order to verify numerical results. Moreover, the logarithmic corrections to scaling \cite{Cassanello1996} for $r=1$-pseudogap model is not considered in this work. However, these limitations open several interesting directions for future investigation. First, one might consider low-temperature approach, in particular the numerical renormalization group method \cite{Bulla2008}, to treat the problem completely. Second, one might employ previous knowledge of perturbation theories \cite{Dai2005,Yosida1975,Horvatic1980,Tong2015} for analytic results in the asymtotic limits, which may justify our numerical results. Other questions of how the logarithmic corrections to scaling enter the numerical results, how the results change when using a more realistic $d$-orbital magnetic impurity, or about the effect of substrates to the graphene layer and the impurity \cite{Ren2014,Donati2014} are also interesting open challenges.

\section*{Acknowledgments}
We thank T. A. Costi for helpful discussion. This research is funded by Vietnam National Foundation for Science and Technology Development (NAFOSTED) under grant number 103.01-2018.12. We also acknowledge the support for the allocation of computing time at J\"ulich Supercomputing Centre. Portion of our calculation as well as data post-processing tasks has been performed using the computer cluster of Phenikaa Institute for Advanced Study.

\appendix
\section{Particle-hole symmetry\label{app:ph_symmetry}}

Previous works on pseudogap Kondo problems \cite{Gonzalez-Buxton1998,Glossop2011,Pixley2012,Fritz2013} do not consider specific lattice of the host materials. Instead, the host material is represented by a ``pseudogap'' DOS, and the hybridization between the impurity and the host material is assumed isotropic. Thus there is no lattice symmetry embedded in these calculations.

In this work, the honeycomb lattice is explicitly included in our calculations. The position of the impurity on the lattice plays an important role in deciding the symmetry of the whole system, and consequently affecting the Kondo physics. In this appendix, we investigate the conditions for the particle-hole symmetry, which controls the form of the hybridization function. The particle-hole symmetry can be understood conceptually as the invariance of the overall Hamiltonian when particles are transformed into holes. We consider a general particle-hole transformation $c_{k\alpha} = U_{\alpha\beta}(k)a^\dagger_{-k\beta}$ ($\hat{U}(k)$ is a $2\times 2$ unitary matrix) \cite{El-Batanouny2020,Franz2019} and apply it on each part of the full Hamiltonian, assuming the impurity is an $s$-wave one. In a quadratic Hamiltonian (such as Eq.~\eqref{eq:h_bath} of the Hamiltonian for the host material)
\[
H = \sum_k h_{\alpha\beta}(k) c^\dagger_{k\alpha} c_{k\beta},
\]
the conditions for the particle-hole symmetry of this Hamiltonian are
\begin{align}
\label{eq:ph_quad_conditions}
\begin{split}
 U^\dagger(k) h(k) U(k) =& -h^*(-k), \\
 \mathrm{Tr[h(k)]} =& 0.
\end{split}
\end{align}
The corresponding transformation matrix for lattice electron creation/annihilation operators is $U=e^{i\phi_k} \sigma_z$ where $\phi_k$ is an arbitrary $k$-dependent phase and $\sigma_z$ is the Pauli matrix along the $z$ direction \cite{Franz2019}, i.e.
\begin{align}
\label{eq:ph_transform1}
\begin{split}
c_{Ak\sigma} &= e^{i\phi_k}a^\dagger_{A,-k\sigma}\\
c_{Bk\sigma} &= -e^{i\phi_k}a^\dagger_{B,-k\sigma}.
\end{split}
\end{align}
Thus from Eq.~\eqref{eq:ph_quad_conditions}, the necessary condition for the particle-hole symmetry is that only the nearest-neighbor hopping is taken into account in Eq.~\eqref{eq:h_bath} and the system is at the charge neutral state ($\mu = 0$).

Considering the impurity local Hamiltonian $H_{loc}$ in Eq.~\eqref{eq:h_loc}, the particle-hole transformation for impurity electron creation/annihilation operators has the form
\begin{equation}
\label{eq:ph_transform2}
d_\sigma = \pm e^{i\phi_k} b^\dagger_\sigma.
\end{equation}
Notice that we choose the same phase $\phi_k$ for both host electron and impurity electron transformation, which is necessary to maintain the particle-hole symmetry if there is, while the sign can be chosen arbitrarily. Applying this transformation on $H_{loc}$, it becomes
\begin{equation}
 H_{loc} = 2\epsilon_{d} + U - \sum_{\sigma} (\epsilon_{d}+U) b^\dagger_\sigma b_\sigma + U n_{b\uparrow} n_{b\downarrow}.
\end{equation}
Thus $H_{loc}$ remains the same form when $\epsilon_d = -U/2$.

Finally, we apply Eqs.~\eqref{eq:ph_transform1} and \eqref{eq:ph_transform2} to $H_{hyb}$ in Eq.~\eqref{eq:h_hyb} where the impurity only hybridizes with the nearest neighbor atoms of the lattice. Without the loss of generality, we choose the sign to be $(+)$ in Eq.~\eqref{eq:ph_transform2}, for single impurity, $H_{hyb}$ becomes
\begin{equation}
\label{eq:h_hyb_transformed}
\begin{split}
 H_{hyb} &= \sum_{k\sigma} (V_{Ak} a_{A,-k\sigma} b^\dagger_\sigma - V_{Bk} a_{B,-k\sigma} b^\dagger_\sigma + h.c.) \\
 &= \sum_{k\sigma} (-V^*_{Ak} b^\dagger_\sigma a_{A,k\sigma} + V^*_{Bk} b^\dagger_\sigma a_{B,k\sigma} + h.c.).
\end{split}
\end{equation}
One sees that the original form of $H_{hyb}$ is no longer maintained when the impurity hybridizes with atoms of both sublattices, i.e. the particle-hole symmetry does not exist, no matter how we choose the sign in Eq.~\eqref{eq:ph_transform2}. The particle-hole symmetry is maintained if the impurity hybridizes with atoms of only one sublattice (with an appropriate choice of the sign in Eq.~\eqref{eq:ph_transform2}). For example, in Eq.\eqref{eq:h_hyb_transformed}, if $V_{Ak}$ vanishes, the particle-hole symmetry is conserved.

Therefore, the requirements for particle-hole symmetry in this model are: (1) the hopping of conduction electrons to sites further than the nearest neighbor sites in the host is forbidded, (2) the impurity energy level is at half-filling $\epsilon_d = -U/2$ and (3) there is only hybridization between the impurity and sites of the same sublattice. Among the three positions $t$, $b$ and $h$ as in Fig.~\ref{fig:imp_graphene}(a), impurity at the $t$ site is more likely to exhibit particle-hole symmetry as the impurity is hybridized mostly to the site underneath, while impurity at $b$ and $h$ sites is hybridized with sites of both sublattices. The result is depicted in the hybridization spectrum in Fig.~\ref{fig:hyb_dos}. At the $t$ site, the spectrum is symmetric around $\omega = 0$ and the particle-hole symmetry is reached when $\epsilon_d = -\frac{U}{2}$. At $b$ and $h$ sites, condition (3) is not satisfied, exhibiting asymmetric hybridization spectrum, there is no particle-hole symmetric point.

Defining the degree of asymmetry by the ratio between the weights of the hybridization spectrum below and above the Fermi level, one sees that this ratio is unity for impurity at the $t$ site, but increases as the impurity changes from $b$ to $h$ site. It characterizes how the impurity hybridizes with the host material. At the $t$ site, the impurity hybridizes with an atom of only one sublattice, the spectrum is symmetric. At the $b$ site, it hybridizes with two atoms belonging to two sublattices, the degree of asymmetry increases. At the $h$ site, it hybridizes with six atoms, the degree of asymmetry is the largest one. The dependence of the degree of asymmetry in the hybridization spectrum on how the impurity hybridizes with lattice atoms reflects condition (3) for the particle-hole symmetry in this model.

\section{Occupancy and susceptibility\label{app:occ_susc}}

\begin{figure}
\includegraphics[width=\columnwidth]{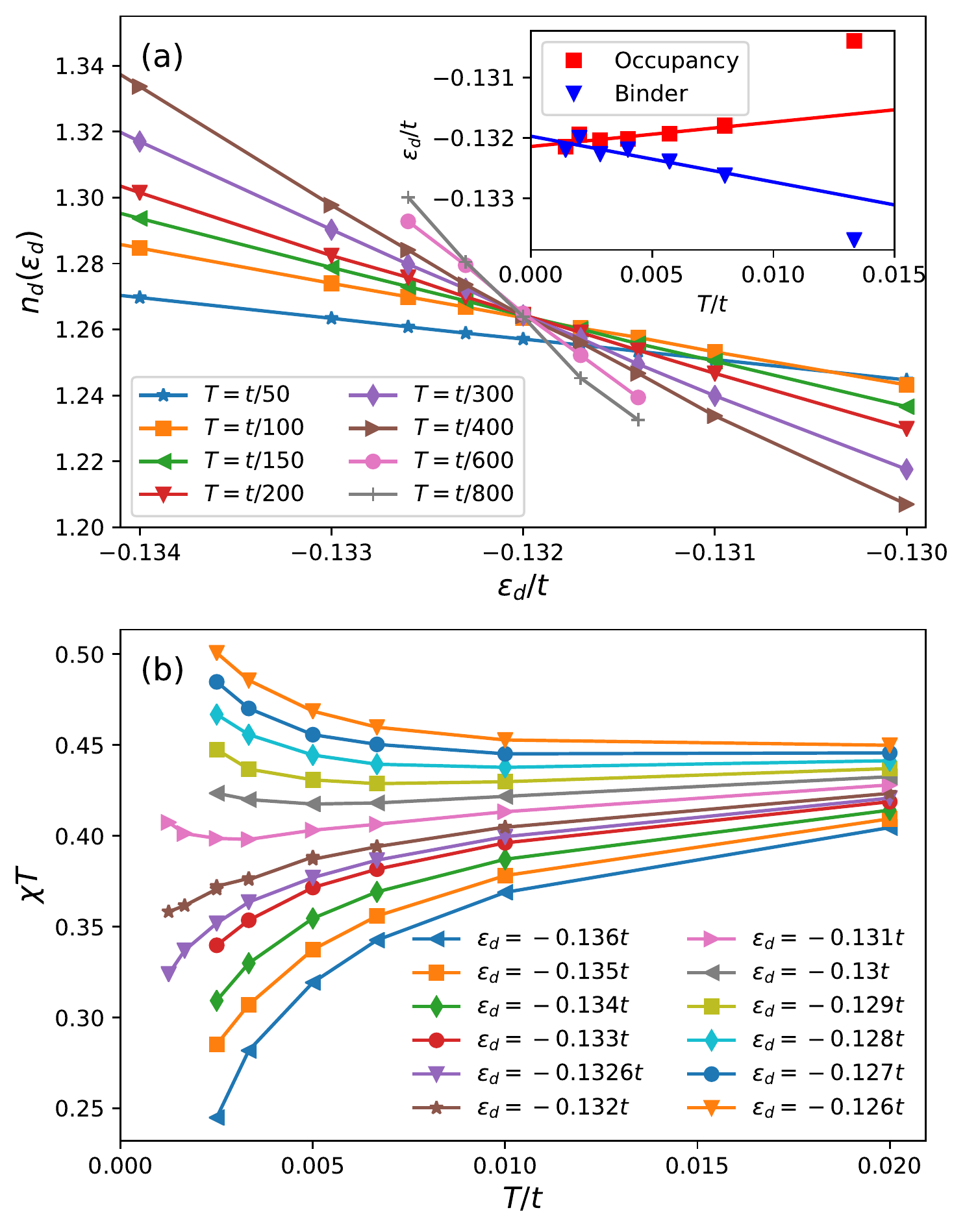}
\caption{\label{fig:occ_chi} CT-HYB measurements for the case $U=0.15t$ and $v_0=0.5t$ (the case of Fig.~\ref{fig:benchmark}) at different temperatures $T$ and impurity energy levels $\epsilon_d$. (a) Occupancy $n_d$ vs. $\epsilon_d$: each line corresponds to a fixed temperature $T$. The dot symbols mark the points undergoing CT-HYB simulations. Inset: comparison of two types of extrapolation for critical point $\epsilon_d^c$ using the Binder parameter ($\epsilon_d^c = -0.13197(9)t$) and the occupancy ($\epsilon_d^c = -0.13214(6)t$). (b) Impurity spin susceptibility (multiplied by $T$) vs. temperature $T$. Each line corresponds to a fixed impurity level $\epsilon_d$.}
\end{figure} 

Critical point for the quantum phase transition between the free local moment and the Kondo strong coupling phases is determined using the Binder analysis \cite{Binder1981,Glossop2011, Pixley2012} as presented in Sec.~\ref{sec:model_methods}. When varying the temperature and the impurity energy level in order to observe the crossing points of Binder parameter lines, we also observe the behaviors of the impurity occupancy and the impurity spin susceptibility, which supports that the point obtained from the Binder analysis is indeed a critical point.

Figure~\ref{fig:occ_chi}(a) shows the electron occupancy of the impurity vs. $\epsilon_d$, which resembles the Binder parameter plotted in Fig.~\ref{fig:benchmark}. At $\epsilon_d^c = -0.13197(9)t$ (determined by the Binder analysis), the occupancy is nearly fixed at $n_d\sim 1.263$. The way the impurity occupancy converges depends very much on the DOS around the Fermi level. Assuming the Fermi-Dirac distribution applied for the quasiparticles, at temperature $T$, the impurity occupancy has more contribution of the DOS for energy in the range $0 < \omega < T$ and less contribution of the DOS for energy in the range $-T < \omega < 0$. When $T$ decreases, the way the impurity occupancy increases or decreases depends on whether the DOS in the range $0 < \omega < T$ is larger or smaller than that in the range $-T < \omega < 0$. If it is in the Kondo phase ($\epsilon_d < \epsilon_d^c$), the Kondo resonance peak is developed at the Fermi level, thus the impurity occupancy increases as $T$ decreases. In contrast, if it is in the free local-moment phase ($\epsilon_d > \epsilon_d^c$), the Kondo peak is suppressed, the DOS for $0 < \omega < T$ is comparable with that around the Fermi level, thus the impurity occupancy decreases as $T$ decreases. The occupancy at criticality becomes a fixed point and can be extrapolated in terms of the crossing points similar to the Binder analysis.

The inset of Fig.~\ref{fig:occ_chi} compares the extrapolation of $\epsilon_d^c$ using the occupancy and the Binder parameter methods. The two fiting lines corresponding to the occupancy and the Binder parameter have intercepts rather close to each other. However, the convergence of the critical impurity occupancy depends much on the hybridization strength $v_0$. For large $v_0$, it requires simulations at low temperatures to observe $n_d$ converging to a critical value. On the other hand, Binder analysis is an well-established theoretical framework to determine critical points for finite-size simulations \cite{Binder1981,Pixley2012}, which is able to exhibit linear convergence at larger temperatures. Therefore, Binder analysis is the method of choice to determine the critical point in this work. Nevertheless, the demonstration of the impurity occupancy as a fixed point in Fig.~\ref{fig:occ_chi} justifies the Binder analysis approach.

The QCP is also confirmed by observing the evolution of the impurity local spin susceptibility $\chi_s$ as the impurity level crosses the critical point $\epsilon_d^c$. The local spin susceptibility of the impurity is calculated using the spin-spin correlation function measured in the CT-HYB simulation
\begin{equation}
 \chi_s = \int_0^\beta d\tau \langle \sigma_z(\tau) \sigma_z(0)\rangle,
\end{equation}
where $\sigma_z = n_{d\uparrow} - n_{d\downarrow}$. The product $\chi_s T$ is a useful criteria to see if the local moment is quenched by conduction electrons of the host \cite{Wilson1975,Bulla2008}. In this aspect, one observes the bending of $\chi T$ curve to predict the phase transition. As in Figure~\ref{fig:occ_chi}(b), for $\epsilon_d < \epsilon_d^c$ the curves bend down to zero, implying the local moment is screened \cite{Krishna-murthy1980a,Krishna-murthy1980b,Bulla2008}; whereas for $\epsilon_d > \epsilon_d^c$, the curves go up to larger values, suggesting the free fluctuation of the local moment. Using $\epsilon_d^c = -0.13197(9)t$ obtained from Binder analysis, the behaviors of $\chi_s T$ well above and below $\epsilon_d^c$ indeed follow our description. At $\epsilon_d$ close to $\epsilon_d^c$, it is not easy to determine if this $\epsilon_d$ belongs to the Kondo screening phase based on $\chi T$, unless one conducts simulations at low temperatures. In Fig.~\ref{fig:occ_chi}(b), detailed calculations of $\chi_s T$ at $T = t/800$ shows that the critical point is indeed in the range of $\epsilon_d$ from $-0.132t$ to $-0.131t$, compatible with the prediction from Binder analysis. However, the calculations of spin susceptibility at low temperature ($T = t/600$ and $t/800$) are computationally demanding, using $\chi_s T$ is thus not an optimized approach to determine the QCP, especially for high-temperature methods such as CT-HYB. Nevertheless, the two distinguished behaviors of $\chi_s T$ above and below the critical $\epsilon_d^c$ provide another evidence that the quantum phase transition occurs around this value.

\bibliography{references}

\begin{thebibliography}{59}%
\makeatletter
\providecommand \@ifxundefined [1]{%
 \@ifx{#1\undefined}
}%
\providecommand \@ifnum [1]{%
 \ifnum #1\expandafter \@firstoftwo
 \else \expandafter \@secondoftwo
 \fi
}%
\providecommand \@ifx [1]{%
 \ifx #1\expandafter \@firstoftwo
 \else \expandafter \@secondoftwo
 \fi
}%
\providecommand \natexlab [1]{#1}%
\providecommand \enquote  [1]{``#1''}%
\providecommand \bibnamefont  [1]{#1}%
\providecommand \bibfnamefont [1]{#1}%
\providecommand \citenamefont [1]{#1}%
\providecommand \href@noop [0]{\@secondoftwo}%
\providecommand \href [0]{\begingroup \@sanitize@url \@href}%
\providecommand \@href[1]{\@@startlink{#1}\@@href}%
\providecommand \@@href[1]{\endgroup#1\@@endlink}%
\providecommand \@sanitize@url [0]{\catcode `\\12\catcode `\$12\catcode
  `\&12\catcode `\#12\catcode `\^12\catcode `\_12\catcode `\%12\relax}%
\providecommand \@@startlink[1]{}%
\providecommand \@@endlink[0]{}%
\providecommand \url  [0]{\begingroup\@sanitize@url \@url }%
\providecommand \@url [1]{\endgroup\@href {#1}{\urlprefix }}%
\providecommand \urlprefix  [0]{URL }%
\providecommand \Eprint [0]{\href }%
\providecommand \doibase [0]{http://dx.doi.org/}%
\providecommand \selectlanguage [0]{\@gobble}%
\providecommand \bibinfo  [0]{\@secondoftwo}%
\providecommand \bibfield  [0]{\@secondoftwo}%
\providecommand \translation [1]{[#1]}%
\providecommand \BibitemOpen [0]{}%
\providecommand \bibitemStop [0]{}%
\providecommand \bibitemNoStop [0]{.\EOS\space}%
\providecommand \EOS [0]{\spacefactor3000\relax}%
\providecommand \BibitemShut  [1]{\csname bibitem#1\endcsname}%
\let\auto@bib@innerbib\@empty
\bibitem [{\citenamefont {Hewson}(1997)}]{Hewson1997}%
  \BibitemOpen
  \bibfield  {author} {\bibinfo {author} {\bibfnamefont {Alexander~Cyril}\
  \bibnamefont {Hewson}},\ }\href {https://doi.org/10.1017/CBO9780511470752}
  {\emph {\bibinfo {title} {The Kondo problem to heavy fermions}}},\
  Vol.~\bibinfo {volume} {2}\ (\bibinfo  {publisher} {Cambridge university
  press},\ \bibinfo {year} {1997})\BibitemShut {NoStop}%
\bibitem [{\citenamefont {Kondo}(1964)}]{Kondo1964}%
  \BibitemOpen
  \bibfield  {author} {\bibinfo {author} {\bibfnamefont {Jun}\ \bibnamefont
  {Kondo}},\ }\bibfield  {title} {\enquote {\bibinfo {title} {{Resistance
  Minimum in Dilute Magnetic Alloys}},}\ }\href {\doibase 10.1143/PTP.32.37}
  {\bibfield  {journal} {\bibinfo  {journal} {Progress of Theoretical Physics}\
  }\textbf {\bibinfo {volume} {32}},\ \bibinfo {pages} {37--49} (\bibinfo
  {year} {1964})}\BibitemShut {NoStop}%
\bibitem [{\citenamefont {Wilson}(1975)}]{Wilson1975}%
  \BibitemOpen
  \bibfield  {author} {\bibinfo {author} {\bibfnamefont {Kenneth~G.}\
  \bibnamefont {Wilson}},\ }\bibfield  {title} {\enquote {\bibinfo {title} {The
  renormalization group: Critical phenomena and the kondo problem},}\ }\href
  {\doibase 10.1103/RevModPhys.47.773} {\bibfield  {journal} {\bibinfo
  {journal} {Rev. Mod. Phys.}\ }\textbf {\bibinfo {volume} {47}},\ \bibinfo
  {pages} {773--840} (\bibinfo {year} {1975})}\BibitemShut {NoStop}%
\bibitem [{\citenamefont {Anderson}(1961)}]{Anderson1961}%
  \BibitemOpen
  \bibfield  {author} {\bibinfo {author} {\bibfnamefont {P.~W.}\ \bibnamefont
  {Anderson}},\ }\bibfield  {title} {\enquote {\bibinfo {title} {Localized
  magnetic states in metals},}\ }\href {\doibase 10.1103/PhysRev.124.41}
  {\bibfield  {journal} {\bibinfo  {journal} {Phys. Rev.}\ }\textbf {\bibinfo
  {volume} {124}},\ \bibinfo {pages} {41--53} (\bibinfo {year}
  {1961})}\BibitemShut {NoStop}%
\bibitem [{\citenamefont {Bulla}\ \emph {et~al.}(2008)\citenamefont {Bulla},
  \citenamefont {Costi},\ and\ \citenamefont {Pruschke}}]{Bulla2008}%
  \BibitemOpen
  \bibfield  {author} {\bibinfo {author} {\bibfnamefont {Ralf}\ \bibnamefont
  {Bulla}}, \bibinfo {author} {\bibfnamefont {Theo~A.}\ \bibnamefont {Costi}},
  \ and\ \bibinfo {author} {\bibfnamefont {Thomas}\ \bibnamefont {Pruschke}},\
  }\bibfield  {title} {\enquote {\bibinfo {title} {Numerical renormalization
  group method for quantum impurity systems},}\ }\href {\doibase
  10.1103/RevModPhys.80.395} {\bibfield  {journal} {\bibinfo  {journal} {Rev.
  Mod. Phys.}\ }\textbf {\bibinfo {volume} {80}},\ \bibinfo {pages} {395--450}
  (\bibinfo {year} {2008})}\BibitemShut {NoStop}%
\bibitem [{\citenamefont {Doniach}(1977)}]{Doniach1977}%
  \BibitemOpen
  \bibfield  {author} {\bibinfo {author} {\bibfnamefont {S.}~\bibnamefont
  {Doniach}},\ }\bibfield  {title} {\enquote {\bibinfo {title} {The kondo
  lattice and weak antiferromagnetism},}\ }\href {\doibase
  https://doi.org/10.1016/0378-4363(77)90190-5} {\bibfield  {journal} {\bibinfo
   {journal} {Physica B+C}\ }\textbf {\bibinfo {volume} {91}},\ \bibinfo
  {pages} {231--234} (\bibinfo {year} {1977})}\BibitemShut {NoStop}%
\bibitem [{\citenamefont {L\"ohneysen}\ \emph {et~al.}(2007)\citenamefont
  {L\"ohneysen}, \citenamefont {Rosch}, \citenamefont {Vojta},\ and\
  \citenamefont {W\"olfle}}]{Loehneysen2007}%
  \BibitemOpen
  \bibfield  {author} {\bibinfo {author} {\bibfnamefont {Hilbert~v.}\
  \bibnamefont {L\"ohneysen}}, \bibinfo {author} {\bibfnamefont {Achim}\
  \bibnamefont {Rosch}}, \bibinfo {author} {\bibfnamefont {Matthias}\
  \bibnamefont {Vojta}}, \ and\ \bibinfo {author} {\bibfnamefont {Peter}\
  \bibnamefont {W\"olfle}},\ }\bibfield  {title} {\enquote {\bibinfo {title}
  {Fermi-liquid instabilities at magnetic quantum phase transitions},}\ }\href
  {\doibase 10.1103/RevModPhys.79.1015} {\bibfield  {journal} {\bibinfo
  {journal} {Rev. Mod. Phys.}\ }\textbf {\bibinfo {volume} {79}},\ \bibinfo
  {pages} {1015--1075} (\bibinfo {year} {2007})}\BibitemShut {NoStop}%
\bibitem [{\citenamefont {Gegenwart}\ \emph {et~al.}(2008)\citenamefont
  {Gegenwart}, \citenamefont {Si},\ and\ \citenamefont
  {Steglich}}]{Gegenwart2008}%
  \BibitemOpen
  \bibfield  {author} {\bibinfo {author} {\bibfnamefont {Philipp}\ \bibnamefont
  {Gegenwart}}, \bibinfo {author} {\bibfnamefont {Qimiao}\ \bibnamefont {Si}},
  \ and\ \bibinfo {author} {\bibfnamefont {Frank}\ \bibnamefont {Steglich}},\
  }\bibfield  {title} {\enquote {\bibinfo {title} {Quantum criticality in
  heavy-fermion metals},}\ }\href@noop {} {\bibfield  {journal} {\bibinfo
  {journal} {nature physics}\ }\textbf {\bibinfo {volume} {4}},\ \bibinfo
  {pages} {186--197} (\bibinfo {year} {2008})}\BibitemShut {NoStop}%
\bibitem [{\citenamefont {Georges}\ and\ \citenamefont
  {Kotliar}(1992)}]{Georges1992}%
  \BibitemOpen
  \bibfield  {author} {\bibinfo {author} {\bibfnamefont {Antoine}\ \bibnamefont
  {Georges}}\ and\ \bibinfo {author} {\bibfnamefont {Gabriel}\ \bibnamefont
  {Kotliar}},\ }\bibfield  {title} {\enquote {\bibinfo {title} {Hubbard model
  in infinite dimensions},}\ }\href {\doibase 10.1103/PhysRevB.45.6479}
  {\bibfield  {journal} {\bibinfo  {journal} {Phys. Rev. B}\ }\textbf {\bibinfo
  {volume} {45}},\ \bibinfo {pages} {6479--6483} (\bibinfo {year}
  {1992})}\BibitemShut {NoStop}%
\bibitem [{\citenamefont {Georges}\ \emph {et~al.}(1996)\citenamefont
  {Georges}, \citenamefont {Kotliar}, \citenamefont {Krauth},\ and\
  \citenamefont {Rozenberg}}]{Georges1996}%
  \BibitemOpen
  \bibfield  {author} {\bibinfo {author} {\bibfnamefont {Antoine}\ \bibnamefont
  {Georges}}, \bibinfo {author} {\bibfnamefont {Gabriel}\ \bibnamefont
  {Kotliar}}, \bibinfo {author} {\bibfnamefont {Werner}\ \bibnamefont
  {Krauth}}, \ and\ \bibinfo {author} {\bibfnamefont {Marcelo~J.}\ \bibnamefont
  {Rozenberg}},\ }\bibfield  {title} {\enquote {\bibinfo {title} {Dynamical
  mean-field theory of strongly correlated fermion systems and the limit of
  infinite dimensions},}\ }\href {\doibase 10.1103/RevModPhys.68.13} {\bibfield
   {journal} {\bibinfo  {journal} {Rev. Mod. Phys.}\ }\textbf {\bibinfo
  {volume} {68}},\ \bibinfo {pages} {13--125} (\bibinfo {year}
  {1996})}\BibitemShut {NoStop}%
\bibitem [{\citenamefont {Georges}\ \emph {et~al.}(2013)\citenamefont
  {Georges}, \citenamefont {Medici},\ and\ \citenamefont
  {Mravlje}}]{Georges2013}%
  \BibitemOpen
  \bibfield  {author} {\bibinfo {author} {\bibfnamefont {Antoine}\ \bibnamefont
  {Georges}}, \bibinfo {author} {\bibfnamefont {Luca~de'}\ \bibnamefont
  {Medici}}, \ and\ \bibinfo {author} {\bibfnamefont {Jernej}\ \bibnamefont
  {Mravlje}},\ }\bibfield  {title} {\enquote {\bibinfo {title} {Strong
  correlations from hund’s coupling},}\ }\href {\doibase
  10.1146/annurev-conmatphys-020911-125045} {\bibfield  {journal} {\bibinfo
  {journal} {Annual Review of Condensed Matter Physics}\ }\textbf {\bibinfo
  {volume} {4}},\ \bibinfo {pages} {137--178} (\bibinfo {year}
  {2013})}\BibitemShut {NoStop}%
\bibitem [{\citenamefont {Gull}\ \emph {et~al.}(2011)\citenamefont {Gull},
  \citenamefont {Millis}, \citenamefont {Lichtenstein}, \citenamefont
  {Rubtsov}, \citenamefont {Troyer},\ and\ \citenamefont {Werner}}]{Gull2011}%
  \BibitemOpen
  \bibfield  {author} {\bibinfo {author} {\bibfnamefont {Emanuel}\ \bibnamefont
  {Gull}}, \bibinfo {author} {\bibfnamefont {Andrew~J.}\ \bibnamefont
  {Millis}}, \bibinfo {author} {\bibfnamefont {Alexander~I.}\ \bibnamefont
  {Lichtenstein}}, \bibinfo {author} {\bibfnamefont {Alexey~N.}\ \bibnamefont
  {Rubtsov}}, \bibinfo {author} {\bibfnamefont {Matthias}\ \bibnamefont
  {Troyer}}, \ and\ \bibinfo {author} {\bibfnamefont {Philipp}\ \bibnamefont
  {Werner}},\ }\bibfield  {title} {\enquote {\bibinfo {title} {Continuous-time
  monte carlo methods for quantum impurity models},}\ }\href {\doibase
  10.1103/RevModPhys.83.349} {\bibfield  {journal} {\bibinfo  {journal} {Rev.
  Mod. Phys.}\ }\textbf {\bibinfo {volume} {83}},\ \bibinfo {pages} {349--404}
  (\bibinfo {year} {2011})}\BibitemShut {NoStop}%
\bibitem [{\citenamefont {Kouwenhoven}\ and\ \citenamefont
  {Glazman}(2001)}]{Kouwenhoven2001}%
  \BibitemOpen
  \bibfield  {author} {\bibinfo {author} {\bibfnamefont {Leo}\ \bibnamefont
  {Kouwenhoven}}\ and\ \bibinfo {author} {\bibfnamefont {Leonid}\ \bibnamefont
  {Glazman}},\ }\bibfield  {title} {\enquote {\bibinfo {title} {Revival of the
  kondo effect},}\ }\href {\doibase 10.1088/2058-7058/14/1/28} {\bibfield
  {journal} {\bibinfo  {journal} {Physics World}\ }\textbf {\bibinfo {volume}
  {14}},\ \bibinfo {pages} {33--38} (\bibinfo {year} {2001})}\BibitemShut
  {NoStop}%
\bibitem [{\citenamefont {Goldhaber-Gordon}\ \emph {et~al.}(1998)\citenamefont
  {Goldhaber-Gordon}, \citenamefont {Shtrikman}, \citenamefont {Mahalu},
  \citenamefont {Abusch-Magder}, \citenamefont {Meirav},\ and\ \citenamefont
  {Kastner}}]{GoldhaberGordon1998}%
  \BibitemOpen
  \bibfield  {author} {\bibinfo {author} {\bibfnamefont {David}\ \bibnamefont
  {Goldhaber-Gordon}}, \bibinfo {author} {\bibfnamefont {Hadas}\ \bibnamefont
  {Shtrikman}}, \bibinfo {author} {\bibfnamefont {D}~\bibnamefont {Mahalu}},
  \bibinfo {author} {\bibfnamefont {David}\ \bibnamefont {Abusch-Magder}},
  \bibinfo {author} {\bibfnamefont {U}~\bibnamefont {Meirav}}, \ and\ \bibinfo
  {author} {\bibfnamefont {MA}~\bibnamefont {Kastner}},\ }\bibfield  {title}
  {\enquote {\bibinfo {title} {Kondo effect in a single-electron transistor},}\
  }\href@noop {} {\bibfield  {journal} {\bibinfo  {journal} {Nature}\ }\textbf
  {\bibinfo {volume} {391}},\ \bibinfo {pages} {156--159} (\bibinfo {year}
  {1998})}\BibitemShut {NoStop}%
\bibitem [{\citenamefont {Sasaki}\ \emph {et~al.}(2000)\citenamefont {Sasaki},
  \citenamefont {De~Franceschi}, \citenamefont {Elzerman}, \citenamefont
  {Van~der Wiel}, \citenamefont {Eto}, \citenamefont {Tarucha},\ and\
  \citenamefont {Kouwenhoven}}]{Sasaki2000}%
  \BibitemOpen
  \bibfield  {author} {\bibinfo {author} {\bibfnamefont {S}~\bibnamefont
  {Sasaki}}, \bibinfo {author} {\bibfnamefont {S}~\bibnamefont
  {De~Franceschi}}, \bibinfo {author} {\bibfnamefont {JM}~\bibnamefont
  {Elzerman}}, \bibinfo {author} {\bibfnamefont {WG}~\bibnamefont {Van~der
  Wiel}}, \bibinfo {author} {\bibfnamefont {Mikio}\ \bibnamefont {Eto}},
  \bibinfo {author} {\bibfnamefont {S}~\bibnamefont {Tarucha}}, \ and\ \bibinfo
  {author} {\bibfnamefont {LP}~\bibnamefont {Kouwenhoven}},\ }\bibfield
  {title} {\enquote {\bibinfo {title} {Kondo effect in an integer-spin quantum
  dot},}\ }\href@noop {} {\bibfield  {journal} {\bibinfo  {journal} {Nature}\
  }\textbf {\bibinfo {volume} {405}},\ \bibinfo {pages} {764--767} (\bibinfo
  {year} {2000})}\BibitemShut {NoStop}%
\bibitem [{\citenamefont {Crommie}\ \emph
  {et~al.}(1993{\natexlab{a}})\citenamefont {Crommie}, \citenamefont {Lutz},\
  and\ \citenamefont {Eigler}}]{Crommie1993a}%
  \BibitemOpen
  \bibfield  {author} {\bibinfo {author} {\bibfnamefont {M.~F.}\ \bibnamefont
  {Crommie}}, \bibinfo {author} {\bibfnamefont {C.~P.}\ \bibnamefont {Lutz}}, \
  and\ \bibinfo {author} {\bibfnamefont {D.~M.}\ \bibnamefont {Eigler}},\
  }\bibfield  {title} {\enquote {\bibinfo {title} {Spectroscopy of a single
  adsorbed atom},}\ }\href {\doibase 10.1103/PhysRevB.48.2851} {\bibfield
  {journal} {\bibinfo  {journal} {Phys. Rev. B}\ }\textbf {\bibinfo {volume}
  {48}},\ \bibinfo {pages} {2851--2854} (\bibinfo {year}
  {1993}{\natexlab{a}})}\BibitemShut {NoStop}%
\bibitem [{\citenamefont {Crommie}\ \emph
  {et~al.}(1993{\natexlab{b}})\citenamefont {Crommie}, \citenamefont {Lutz},\
  and\ \citenamefont {Eigler}}]{Crommie1993b}%
  \BibitemOpen
  \bibfield  {author} {\bibinfo {author} {\bibfnamefont {MF}~\bibnamefont
  {Crommie}}, \bibinfo {author} {\bibfnamefont {Ch~P}\ \bibnamefont {Lutz}}, \
  and\ \bibinfo {author} {\bibfnamefont {DM}~\bibnamefont {Eigler}},\
  }\bibfield  {title} {\enquote {\bibinfo {title} {Imaging standing waves in a
  two-dimensional electron gas},}\ }\href@noop {} {\bibfield  {journal}
  {\bibinfo  {journal} {Nature}\ }\textbf {\bibinfo {volume} {363}},\ \bibinfo
  {pages} {524--527} (\bibinfo {year} {1993}{\natexlab{b}})}\BibitemShut
  {NoStop}%
\bibitem [{\citenamefont {Li}\ \emph {et~al.}(1998)\citenamefont {Li},
  \citenamefont {Schneider}, \citenamefont {Berndt},\ and\ \citenamefont
  {Delley}}]{Li1998}%
  \BibitemOpen
  \bibfield  {author} {\bibinfo {author} {\bibfnamefont {Jiutao}\ \bibnamefont
  {Li}}, \bibinfo {author} {\bibfnamefont {Wolf-Dieter}\ \bibnamefont
  {Schneider}}, \bibinfo {author} {\bibfnamefont {Richard}\ \bibnamefont
  {Berndt}}, \ and\ \bibinfo {author} {\bibfnamefont {Bernard}\ \bibnamefont
  {Delley}},\ }\bibfield  {title} {\enquote {\bibinfo {title} {Kondo scattering
  observed at a single magnetic impurity},}\ }\href {\doibase
  10.1103/PhysRevLett.80.2893} {\bibfield  {journal} {\bibinfo  {journal}
  {Phys. Rev. Lett.}\ }\textbf {\bibinfo {volume} {80}},\ \bibinfo {pages}
  {2893--2896} (\bibinfo {year} {1998})}\BibitemShut {NoStop}%
\bibitem [{\citenamefont {Manoharan}\ \emph {et~al.}(2000)\citenamefont
  {Manoharan}, \citenamefont {Lutz},\ and\ \citenamefont
  {Eigler}}]{Manoharan2000}%
  \BibitemOpen
  \bibfield  {author} {\bibinfo {author} {\bibfnamefont {HC}~\bibnamefont
  {Manoharan}}, \bibinfo {author} {\bibfnamefont {CP}~\bibnamefont {Lutz}}, \
  and\ \bibinfo {author} {\bibfnamefont {DM}~\bibnamefont {Eigler}},\
  }\bibfield  {title} {\enquote {\bibinfo {title} {Quantum mirages formed by
  coherent projection of electronic structure},}\ }\href
  {https://doi.org/10.1038/35000508} {\bibfield  {journal} {\bibinfo  {journal}
  {Nature}\ }\textbf {\bibinfo {volume} {403}},\ \bibinfo {pages} {512--515}
  (\bibinfo {year} {2000})}\BibitemShut {NoStop}%
\bibitem [{\citenamefont {Withoff}\ and\ \citenamefont
  {Fradkin}(1990)}]{Withoff1990}%
  \BibitemOpen
  \bibfield  {author} {\bibinfo {author} {\bibfnamefont {David}\ \bibnamefont
  {Withoff}}\ and\ \bibinfo {author} {\bibfnamefont {Eduardo}\ \bibnamefont
  {Fradkin}},\ }\bibfield  {title} {\enquote {\bibinfo {title} {Phase
  transitions in gapless fermi systems with magnetic impurities},}\ }\href
  {\doibase 10.1103/PhysRevLett.64.1835} {\bibfield  {journal} {\bibinfo
  {journal} {Phys. Rev. Lett.}\ }\textbf {\bibinfo {volume} {64}},\ \bibinfo
  {pages} {1835--1838} (\bibinfo {year} {1990})}\BibitemShut {NoStop}%
\bibitem [{\citenamefont {Gonzalez-Buxton}\ and\ \citenamefont
  {Ingersent}(1998)}]{Gonzalez-Buxton1998}%
  \BibitemOpen
  \bibfield  {author} {\bibinfo {author} {\bibfnamefont {Carlos}\ \bibnamefont
  {Gonzalez-Buxton}}\ and\ \bibinfo {author} {\bibfnamefont {Kevin}\
  \bibnamefont {Ingersent}},\ }\bibfield  {title} {\enquote {\bibinfo {title}
  {Renormalization-group study of anderson and kondo impurities in gapless
  fermi systems},}\ }\href {\doibase 10.1103/PhysRevB.57.14254} {\bibfield
  {journal} {\bibinfo  {journal} {Phys. Rev. B}\ }\textbf {\bibinfo {volume}
  {57}},\ \bibinfo {pages} {14254--14293} (\bibinfo {year} {1998})}\BibitemShut
  {NoStop}%
\bibitem [{\citenamefont {Pixley}\ \emph {et~al.}(2012)\citenamefont {Pixley},
  \citenamefont {Kirchner}, \citenamefont {Ingersent},\ and\ \citenamefont
  {Si}}]{Pixley2012}%
  \BibitemOpen
  \bibfield  {author} {\bibinfo {author} {\bibfnamefont {J.~H.}\ \bibnamefont
  {Pixley}}, \bibinfo {author} {\bibfnamefont {Stefan}\ \bibnamefont
  {Kirchner}}, \bibinfo {author} {\bibfnamefont {Kevin}\ \bibnamefont
  {Ingersent}}, \ and\ \bibinfo {author} {\bibfnamefont {Qimiao}\ \bibnamefont
  {Si}},\ }\bibfield  {title} {\enquote {\bibinfo {title} {Kondo destruction
  and valence fluctuations in an anderson model},}\ }\href {\doibase
  10.1103/PhysRevLett.109.086403} {\bibfield  {journal} {\bibinfo  {journal}
  {Phys. Rev. Lett.}\ }\textbf {\bibinfo {volume} {109}},\ \bibinfo {pages}
  {086403} (\bibinfo {year} {2012})}\BibitemShut {NoStop}%
\bibitem [{\citenamefont {Bulla}\ \emph {et~al.}(1997)\citenamefont {Bulla},
  \citenamefont {Pruschke},\ and\ \citenamefont {Hewson}}]{Bulla1997}%
  \BibitemOpen
  \bibfield  {author} {\bibinfo {author} {\bibfnamefont {R}~\bibnamefont
  {Bulla}}, \bibinfo {author} {\bibfnamefont {Th}~\bibnamefont {Pruschke}}, \
  and\ \bibinfo {author} {\bibfnamefont {A~C}\ \bibnamefont {Hewson}},\
  }\bibfield  {title} {\enquote {\bibinfo {title} {Anderson impurity in
  pseudo-gap fermi systems},}\ }\href {\doibase 10.1088/0953-8984/9/47/014}
  {\bibfield  {journal} {\bibinfo  {journal} {Journal of Physics: Condensed
  Matter}\ }\textbf {\bibinfo {volume} {9}},\ \bibinfo {pages} {10463--10474}
  (\bibinfo {year} {1997})}\BibitemShut {NoStop}%
\bibitem [{\citenamefont {Cassanello}\ and\ \citenamefont
  {Fradkin}(1996)}]{Cassanello1996}%
  \BibitemOpen
  \bibfield  {author} {\bibinfo {author} {\bibfnamefont {Carlos~R.}\
  \bibnamefont {Cassanello}}\ and\ \bibinfo {author} {\bibfnamefont {Eduardo}\
  \bibnamefont {Fradkin}},\ }\bibfield  {title} {\enquote {\bibinfo {title}
  {Kondo effect in flux phases},}\ }\href {\doibase 10.1103/PhysRevB.53.15079}
  {\bibfield  {journal} {\bibinfo  {journal} {Phys. Rev. B}\ }\textbf {\bibinfo
  {volume} {53}},\ \bibinfo {pages} {15079--15094} (\bibinfo {year}
  {1996})}\BibitemShut {NoStop}%
\bibitem [{\citenamefont {Vojta}\ and\ \citenamefont
  {Fritz}(2004)}]{Vojta2004}%
  \BibitemOpen
  \bibfield  {author} {\bibinfo {author} {\bibfnamefont {Matthias}\
  \bibnamefont {Vojta}}\ and\ \bibinfo {author} {\bibfnamefont {Lars}\
  \bibnamefont {Fritz}},\ }\bibfield  {title} {\enquote {\bibinfo {title}
  {Upper critical dimension in a quantum impurity model: Critical theory of the
  asymmetric pseudogap kondo problem},}\ }\href {\doibase
  10.1103/PhysRevB.70.094502} {\bibfield  {journal} {\bibinfo  {journal} {Phys.
  Rev. B}\ }\textbf {\bibinfo {volume} {70}},\ \bibinfo {pages} {094502}
  (\bibinfo {year} {2004})}\BibitemShut {NoStop}%
\bibitem [{\citenamefont {Fritz}\ and\ \citenamefont
  {Vojta}(2004)}]{Fritz2004}%
  \BibitemOpen
  \bibfield  {author} {\bibinfo {author} {\bibfnamefont {Lars}\ \bibnamefont
  {Fritz}}\ and\ \bibinfo {author} {\bibfnamefont {Matthias}\ \bibnamefont
  {Vojta}},\ }\bibfield  {title} {\enquote {\bibinfo {title} {Phase transitions
  in the pseudogap anderson and kondo models: Critical dimensions,
  renormalization group, and local-moment criticality},}\ }\href {\doibase
  10.1103/PhysRevB.70.214427} {\bibfield  {journal} {\bibinfo  {journal} {Phys.
  Rev. B}\ }\textbf {\bibinfo {volume} {70}},\ \bibinfo {pages} {214427}
  (\bibinfo {year} {2004})}\BibitemShut {NoStop}%
\bibitem [{\citenamefont {Si}\ \emph {et~al.}(2001)\citenamefont {Si},
  \citenamefont {Rabello}, \citenamefont {Ingersent},\ and\ \citenamefont
  {Smith}}]{Si2001}%
  \BibitemOpen
  \bibfield  {author} {\bibinfo {author} {\bibfnamefont {Qimiao}\ \bibnamefont
  {Si}}, \bibinfo {author} {\bibfnamefont {Silvio}\ \bibnamefont {Rabello}},
  \bibinfo {author} {\bibfnamefont {Kevin}\ \bibnamefont {Ingersent}}, \ and\
  \bibinfo {author} {\bibfnamefont {J~Lleweilun}\ \bibnamefont {Smith}},\
  }\bibfield  {title} {\enquote {\bibinfo {title} {Locally critical quantum
  phase transitions in strongly correlated metals},}\ }\href@noop {} {\bibfield
   {journal} {\bibinfo  {journal} {Nature}\ }\textbf {\bibinfo {volume}
  {413}},\ \bibinfo {pages} {804--808} (\bibinfo {year} {2001})}\BibitemShut
  {NoStop}%
\bibitem [{\citenamefont {Vojta}\ and\ \citenamefont
  {Bulla}(2001)}]{Vojta2001}%
  \BibitemOpen
  \bibfield  {author} {\bibinfo {author} {\bibfnamefont {Matthias}\
  \bibnamefont {Vojta}}\ and\ \bibinfo {author} {\bibfnamefont {Ralf}\
  \bibnamefont {Bulla}},\ }\bibfield  {title} {\enquote {\bibinfo {title}
  {Kondo effect of impurity moments in $d$-wave superconductors: Quantum phase
  transition and spectral properties},}\ }\href {\doibase
  10.1103/PhysRevB.65.014511} {\bibfield  {journal} {\bibinfo  {journal} {Phys.
  Rev. B}\ }\textbf {\bibinfo {volume} {65}},\ \bibinfo {pages} {014511}
  (\bibinfo {year} {2001})}\BibitemShut {NoStop}%
\bibitem [{\citenamefont {Novoselov}\ \emph {et~al.}(2004)\citenamefont
  {Novoselov}, \citenamefont {Geim}, \citenamefont {Morozov}, \citenamefont
  {Jiang}, \citenamefont {Zhang}, \citenamefont {Dubonos}, \citenamefont
  {Grigorieva},\ and\ \citenamefont {Firsov}}]{Novoselov2004}%
  \BibitemOpen
  \bibfield  {author} {\bibinfo {author} {\bibfnamefont {K.~S.}\ \bibnamefont
  {Novoselov}}, \bibinfo {author} {\bibfnamefont {A.~K.}\ \bibnamefont {Geim}},
  \bibinfo {author} {\bibfnamefont {S.~V.}\ \bibnamefont {Morozov}}, \bibinfo
  {author} {\bibfnamefont {D.}~\bibnamefont {Jiang}}, \bibinfo {author}
  {\bibfnamefont {Y.}~\bibnamefont {Zhang}}, \bibinfo {author} {\bibfnamefont
  {S.~V.}\ \bibnamefont {Dubonos}}, \bibinfo {author} {\bibfnamefont {I.~V.}\
  \bibnamefont {Grigorieva}}, \ and\ \bibinfo {author} {\bibfnamefont {A.~A.}\
  \bibnamefont {Firsov}},\ }\bibfield  {title} {\enquote {\bibinfo {title}
  {Electric field effect in atomically thin carbon films},}\ }\href {\doibase
  10.1126/science.1102896} {\bibfield  {journal} {\bibinfo  {journal}
  {Science}\ }\textbf {\bibinfo {volume} {306}},\ \bibinfo {pages} {666--669}
  (\bibinfo {year} {2004})}\BibitemShut {NoStop}%
\bibitem [{\citenamefont {Mattos}(2009)}]{Mattos2009}%
  \BibitemOpen
  \bibfield  {author} {\bibinfo {author} {\bibfnamefont {Laila Souza~De}\
  \bibnamefont {Mattos}},\ }\emph {\bibinfo {title} {{Correlated electrons
  probed by scanning tunneling microscopy}}},\ \href@noop {} {Ph.D. thesis},\
  \bibinfo  {school} {Stanford University} (\bibinfo {year} {2009})\BibitemShut
  {NoStop}%
\bibitem [{\citenamefont {Jacob}\ \emph {et~al.}(2009)\citenamefont {Jacob},
  \citenamefont {Haule},\ and\ \citenamefont {Kotliar}}]{Jacob2009}%
  \BibitemOpen
  \bibfield  {author} {\bibinfo {author} {\bibfnamefont {D.}~\bibnamefont
  {Jacob}}, \bibinfo {author} {\bibfnamefont {K.}~\bibnamefont {Haule}}, \ and\
  \bibinfo {author} {\bibfnamefont {G.}~\bibnamefont {Kotliar}},\ }\bibfield
  {title} {\enquote {\bibinfo {title} {Kondo effect and conductance of
  nanocontacts with magnetic impurities},}\ }\href {\doibase
  10.1103/PhysRevLett.103.016803} {\bibfield  {journal} {\bibinfo  {journal}
  {Phys. Rev. Lett.}\ }\textbf {\bibinfo {volume} {103}},\ \bibinfo {pages}
  {016803} (\bibinfo {year} {2009})}\BibitemShut {NoStop}%
\bibitem [{\citenamefont {Brar}\ \emph {et~al.}(2011)\citenamefont {Brar},
  \citenamefont {Decker}, \citenamefont {Solowan}, \citenamefont {Wang},
  \citenamefont {Maserati}, \citenamefont {Chan}, \citenamefont {Lee},
  \citenamefont {Girit}, \citenamefont {Zettl}, \citenamefont {Louie} \emph
  {et~al.}}]{Brar2011}%
  \BibitemOpen
  \bibfield  {author} {\bibinfo {author} {\bibfnamefont {Victor~W}\
  \bibnamefont {Brar}}, \bibinfo {author} {\bibfnamefont {R{\'e}gis}\
  \bibnamefont {Decker}}, \bibinfo {author} {\bibfnamefont {Hans-Michael}\
  \bibnamefont {Solowan}}, \bibinfo {author} {\bibfnamefont {Yang}\
  \bibnamefont {Wang}}, \bibinfo {author} {\bibfnamefont {Lorenzo}\
  \bibnamefont {Maserati}}, \bibinfo {author} {\bibfnamefont {Kevin~T}\
  \bibnamefont {Chan}}, \bibinfo {author} {\bibfnamefont {Hoonkyung}\
  \bibnamefont {Lee}}, \bibinfo {author} {\bibfnamefont {{\c{C}}a{\u{g}}lar~O}\
  \bibnamefont {Girit}}, \bibinfo {author} {\bibfnamefont {Alex}\ \bibnamefont
  {Zettl}}, \bibinfo {author} {\bibfnamefont {Steven~G}\ \bibnamefont {Louie}},
   \emph {et~al.},\ }\bibfield  {title} {\enquote {\bibinfo {title}
  {Gate-controlled ionization and screening of cobalt adatoms on a graphene
  surface},}\ }\href {https://doi.org/10.1038/nphys1807} {\bibfield  {journal}
  {\bibinfo  {journal} {Nature Physics}\ }\textbf {\bibinfo {volume} {7}},\
  \bibinfo {pages} {43--47} (\bibinfo {year} {2011})}\BibitemShut {NoStop}%
\bibitem [{\citenamefont {Wang}\ \emph {et~al.}(2012)\citenamefont {Wang},
  \citenamefont {Brar}, \citenamefont {Shytov}, \citenamefont {Wu},
  \citenamefont {Regan}, \citenamefont {Tsai}, \citenamefont {Zettl},
  \citenamefont {Levitov},\ and\ \citenamefont {Crommie}}]{Wang2012}%
  \BibitemOpen
  \bibfield  {author} {\bibinfo {author} {\bibfnamefont {Yang}\ \bibnamefont
  {Wang}}, \bibinfo {author} {\bibfnamefont {Victor~W}\ \bibnamefont {Brar}},
  \bibinfo {author} {\bibfnamefont {Andrey~V}\ \bibnamefont {Shytov}}, \bibinfo
  {author} {\bibfnamefont {Qiong}\ \bibnamefont {Wu}}, \bibinfo {author}
  {\bibfnamefont {William}\ \bibnamefont {Regan}}, \bibinfo {author}
  {\bibfnamefont {Hsin-Zon}\ \bibnamefont {Tsai}}, \bibinfo {author}
  {\bibfnamefont {Alex}\ \bibnamefont {Zettl}}, \bibinfo {author}
  {\bibfnamefont {Leonid~S}\ \bibnamefont {Levitov}}, \ and\ \bibinfo {author}
  {\bibfnamefont {Michael~F}\ \bibnamefont {Crommie}},\ }\bibfield  {title}
  {\enquote {\bibinfo {title} {Mapping dirac quasiparticles near a single
  coulomb impurity on graphene},}\ }\href {https://doi.org/10.1038/nphys2379}
  {\bibfield  {journal} {\bibinfo  {journal} {Nature Physics}\ }\textbf
  {\bibinfo {volume} {8}},\ \bibinfo {pages} {653--657} (\bibinfo {year}
  {2012})}\BibitemShut {NoStop}%
\bibitem [{\citenamefont {Wehling}\ \emph
  {et~al.}(2010{\natexlab{a}})\citenamefont {Wehling}, \citenamefont
  {Balatsky}, \citenamefont {Katsnelson}, \citenamefont {Lichtenstein},\ and\
  \citenamefont {Rosch}}]{Wehling2010a}%
  \BibitemOpen
  \bibfield  {author} {\bibinfo {author} {\bibfnamefont {T.~O.}\ \bibnamefont
  {Wehling}}, \bibinfo {author} {\bibfnamefont {A.~V.}\ \bibnamefont
  {Balatsky}}, \bibinfo {author} {\bibfnamefont {M.~I.}\ \bibnamefont
  {Katsnelson}}, \bibinfo {author} {\bibfnamefont {A.~I.}\ \bibnamefont
  {Lichtenstein}}, \ and\ \bibinfo {author} {\bibfnamefont {A.}~\bibnamefont
  {Rosch}},\ }\bibfield  {title} {\enquote {\bibinfo {title} {Orbitally
  controlled kondo effect of co adatoms on graphene},}\ }\href {\doibase
  10.1103/PhysRevB.81.115427} {\bibfield  {journal} {\bibinfo  {journal} {Phys.
  Rev. B}\ }\textbf {\bibinfo {volume} {81}},\ \bibinfo {pages} {115427}
  (\bibinfo {year} {2010}{\natexlab{a}})}\BibitemShut {NoStop}%
\bibitem [{\citenamefont {Wehling}\ \emph
  {et~al.}(2010{\natexlab{b}})\citenamefont {Wehling}, \citenamefont {Dahal},
  \citenamefont {Lichtenstein}, \citenamefont {Katsnelson}, \citenamefont
  {Manoharan},\ and\ \citenamefont {Balatsky}}]{Wehling2010b}%
  \BibitemOpen
  \bibfield  {author} {\bibinfo {author} {\bibfnamefont {T.~O.}\ \bibnamefont
  {Wehling}}, \bibinfo {author} {\bibfnamefont {H.~P.}\ \bibnamefont {Dahal}},
  \bibinfo {author} {\bibfnamefont {A.~I.}\ \bibnamefont {Lichtenstein}},
  \bibinfo {author} {\bibfnamefont {M.~I.}\ \bibnamefont {Katsnelson}},
  \bibinfo {author} {\bibfnamefont {H.~C.}\ \bibnamefont {Manoharan}}, \ and\
  \bibinfo {author} {\bibfnamefont {A.~V.}\ \bibnamefont {Balatsky}},\
  }\bibfield  {title} {\enquote {\bibinfo {title} {Theory of fano resonances in
  graphene: The influence of orbital and structural symmetries on stm
  spectra},}\ }\href {\doibase 10.1103/PhysRevB.81.085413} {\bibfield
  {journal} {\bibinfo  {journal} {Phys. Rev. B}\ }\textbf {\bibinfo {volume}
  {81}},\ \bibinfo {pages} {085413} (\bibinfo {year}
  {2010}{\natexlab{b}})}\BibitemShut {NoStop}%
\bibitem [{\citenamefont {Eelbo}\ \emph {et~al.}(2013)\citenamefont {Eelbo},
  \citenamefont {Wa\ifmmode~\acute{s}\else \'{s}\fi{}niowska}, \citenamefont
  {Thakur}, \citenamefont {Gyamfi}, \citenamefont {Sachs}, \citenamefont
  {Wehling}, \citenamefont {Forti}, \citenamefont {Starke}, \citenamefont
  {Tieg}, \citenamefont {Lichtenstein},\ and\ \citenamefont
  {Wiesendanger}}]{Eelbo2013}%
  \BibitemOpen
  \bibfield  {author} {\bibinfo {author} {\bibfnamefont {T.}~\bibnamefont
  {Eelbo}}, \bibinfo {author} {\bibfnamefont {M.}~\bibnamefont
  {Wa\ifmmode~\acute{s}\else \'{s}\fi{}niowska}}, \bibinfo {author}
  {\bibfnamefont {P.}~\bibnamefont {Thakur}}, \bibinfo {author} {\bibfnamefont
  {M.}~\bibnamefont {Gyamfi}}, \bibinfo {author} {\bibfnamefont
  {B.}~\bibnamefont {Sachs}}, \bibinfo {author} {\bibfnamefont {T.~O.}\
  \bibnamefont {Wehling}}, \bibinfo {author} {\bibfnamefont {S.}~\bibnamefont
  {Forti}}, \bibinfo {author} {\bibfnamefont {U.}~\bibnamefont {Starke}},
  \bibinfo {author} {\bibfnamefont {C.}~\bibnamefont {Tieg}}, \bibinfo {author}
  {\bibfnamefont {A.~I.}\ \bibnamefont {Lichtenstein}}, \ and\ \bibinfo
  {author} {\bibfnamefont {R.}~\bibnamefont {Wiesendanger}},\ }\bibfield
  {title} {\enquote {\bibinfo {title} {Adatoms and clusters of $3d$ transition
  metals on graphene: Electronic and magnetic configurations},}\ }\href
  {\doibase 10.1103/PhysRevLett.110.136804} {\bibfield  {journal} {\bibinfo
  {journal} {Phys. Rev. Lett.}\ }\textbf {\bibinfo {volume} {110}},\ \bibinfo
  {pages} {136804} (\bibinfo {year} {2013})}\BibitemShut {NoStop}%
\bibitem [{\citenamefont {Virgus}\ \emph {et~al.}(2014)\citenamefont {Virgus},
  \citenamefont {Purwanto}, \citenamefont {Krakauer},\ and\ \citenamefont
  {Zhang}}]{Virgus2014}%
  \BibitemOpen
  \bibfield  {author} {\bibinfo {author} {\bibfnamefont {Yudistira}\
  \bibnamefont {Virgus}}, \bibinfo {author} {\bibfnamefont {Wirawan}\
  \bibnamefont {Purwanto}}, \bibinfo {author} {\bibfnamefont {Henry}\
  \bibnamefont {Krakauer}}, \ and\ \bibinfo {author} {\bibfnamefont {Shiwei}\
  \bibnamefont {Zhang}},\ }\bibfield  {title} {\enquote {\bibinfo {title}
  {Stability, energetics, and magnetic states of cobalt adatoms on graphene},}\
  }\href {\doibase 10.1103/PhysRevLett.113.175502} {\bibfield  {journal}
  {\bibinfo  {journal} {Phys. Rev. Lett.}\ }\textbf {\bibinfo {volume} {113}},\
  \bibinfo {pages} {175502} (\bibinfo {year} {2014})}\BibitemShut {NoStop}%
\bibitem [{\citenamefont {Ren}\ \emph {et~al.}(2014)\citenamefont {Ren},
  \citenamefont {Guo}, \citenamefont {Pan}, \citenamefont {Zhang},
  \citenamefont {Wu}, \citenamefont {Luo}, \citenamefont {Du}, \citenamefont
  {Pantelides},\ and\ \citenamefont {Gao}}]{Ren2014}%
  \BibitemOpen
  \bibfield  {author} {\bibinfo {author} {\bibfnamefont {Jindong}\ \bibnamefont
  {Ren}}, \bibinfo {author} {\bibfnamefont {Haiming}\ \bibnamefont {Guo}},
  \bibinfo {author} {\bibfnamefont {Jinbo}\ \bibnamefont {Pan}}, \bibinfo
  {author} {\bibfnamefont {Yu~Yang}\ \bibnamefont {Zhang}}, \bibinfo {author}
  {\bibfnamefont {Xu}~\bibnamefont {Wu}}, \bibinfo {author} {\bibfnamefont
  {Hong-Gang}\ \bibnamefont {Luo}}, \bibinfo {author} {\bibfnamefont {Shixuan}\
  \bibnamefont {Du}}, \bibinfo {author} {\bibfnamefont {Sokrates~T.}\
  \bibnamefont {Pantelides}}, \ and\ \bibinfo {author} {\bibfnamefont
  {Hong-Jun}\ \bibnamefont {Gao}},\ }\bibfield  {title} {\enquote {\bibinfo
  {title} {Kondo effect of cobalt adatoms on a graphene monolayer controlled by
  substrate-induced ripples},}\ }\href {\doibase 10.1021/nl501425n} {\bibfield
  {journal} {\bibinfo  {journal} {Nano Letters}\ }\textbf {\bibinfo {volume}
  {14}},\ \bibinfo {pages} {4011--4015} (\bibinfo {year} {2014})},\ \bibinfo
  {note} {pMID: 24905855},\ \Eprint
  {http://arxiv.org/abs/https://doi.org/10.1021/nl501425n}
  {https://doi.org/10.1021/nl501425n} \BibitemShut {NoStop}%
\bibitem [{\citenamefont {Donati}\ \emph {et~al.}(2014)\citenamefont {Donati},
  \citenamefont {Gragnaniello}, \citenamefont {Cavallin}, \citenamefont
  {Natterer}, \citenamefont {Dubout}, \citenamefont {Pivetta}, \citenamefont
  {Patthey}, \citenamefont {Dreiser}, \citenamefont {Piamonteze}, \citenamefont
  {Rusponi},\ and\ \citenamefont {Brune}}]{Donati2014}%
  \BibitemOpen
  \bibfield  {author} {\bibinfo {author} {\bibfnamefont {F.}~\bibnamefont
  {Donati}}, \bibinfo {author} {\bibfnamefont {L.}~\bibnamefont
  {Gragnaniello}}, \bibinfo {author} {\bibfnamefont {A.}~\bibnamefont
  {Cavallin}}, \bibinfo {author} {\bibfnamefont {F.~D.}\ \bibnamefont
  {Natterer}}, \bibinfo {author} {\bibfnamefont {Q.}~\bibnamefont {Dubout}},
  \bibinfo {author} {\bibfnamefont {M.}~\bibnamefont {Pivetta}}, \bibinfo
  {author} {\bibfnamefont {F.}~\bibnamefont {Patthey}}, \bibinfo {author}
  {\bibfnamefont {J.}~\bibnamefont {Dreiser}}, \bibinfo {author} {\bibfnamefont
  {C.}~\bibnamefont {Piamonteze}}, \bibinfo {author} {\bibfnamefont
  {S.}~\bibnamefont {Rusponi}}, \ and\ \bibinfo {author} {\bibfnamefont
  {H.}~\bibnamefont {Brune}},\ }\bibfield  {title} {\enquote {\bibinfo {title}
  {Tailoring the magnetism of co atoms on graphene through substrate
  hybridization},}\ }\href {\doibase 10.1103/PhysRevLett.113.177201} {\bibfield
   {journal} {\bibinfo  {journal} {Phys. Rev. Lett.}\ }\textbf {\bibinfo
  {volume} {113}},\ \bibinfo {pages} {177201} (\bibinfo {year}
  {2014})}\BibitemShut {NoStop}%
\bibitem [{\citenamefont {Fritz}\ and\ \citenamefont
  {Vojta}(2013)}]{Fritz2013}%
  \BibitemOpen
  \bibfield  {author} {\bibinfo {author} {\bibfnamefont {Lars}\ \bibnamefont
  {Fritz}}\ and\ \bibinfo {author} {\bibfnamefont {Matthias}\ \bibnamefont
  {Vojta}},\ }\bibfield  {title} {\enquote {\bibinfo {title} {The physics of
  kondo impurities in graphene},}\ }\href {\doibase
  10.1088/0034-4885/76/3/032501} {\bibfield  {journal} {\bibinfo  {journal}
  {Reports on Progress in Physics}\ }\textbf {\bibinfo {volume} {76}},\
  \bibinfo {pages} {032501} (\bibinfo {year} {2013})}\BibitemShut {NoStop}%
\bibitem [{\citenamefont {Binder}(1981)}]{Binder1981}%
  \BibitemOpen
  \bibfield  {author} {\bibinfo {author} {\bibfnamefont {K.}~\bibnamefont
  {Binder}},\ }\bibfield  {title} {\enquote {\bibinfo {title} {Finite size
  scaling analysis of ising model block distribution functions},}\ }\href
  {\doibase 10.1007/BF01293604} {\bibfield  {journal} {\bibinfo  {journal}
  {Zeitschrift f{\"u}r Physik B Condensed Matter}\ }\textbf {\bibinfo {volume}
  {43}},\ \bibinfo {pages} {119--140} (\bibinfo {year} {1981})}\BibitemShut
  {NoStop}%
\bibitem [{\citenamefont {Castro~Neto}\ \emph {et~al.}(2009)\citenamefont
  {Castro~Neto}, \citenamefont {Guinea}, \citenamefont {Peres}, \citenamefont
  {Novoselov},\ and\ \citenamefont {Geim}}]{Neto2009}%
  \BibitemOpen
  \bibfield  {author} {\bibinfo {author} {\bibfnamefont {A.~H.}\ \bibnamefont
  {Castro~Neto}}, \bibinfo {author} {\bibfnamefont {F.}~\bibnamefont {Guinea}},
  \bibinfo {author} {\bibfnamefont {N.~M.~R.}\ \bibnamefont {Peres}}, \bibinfo
  {author} {\bibfnamefont {K.~S.}\ \bibnamefont {Novoselov}}, \ and\ \bibinfo
  {author} {\bibfnamefont {A.~K.}\ \bibnamefont {Geim}},\ }\bibfield  {title}
  {\enquote {\bibinfo {title} {The electronic properties of graphene},}\ }\href
  {\doibase 10.1103/RevModPhys.81.109} {\bibfield  {journal} {\bibinfo
  {journal} {Rev. Mod. Phys.}\ }\textbf {\bibinfo {volume} {81}},\ \bibinfo
  {pages} {109--162} (\bibinfo {year} {2009})}\BibitemShut {NoStop}%
\bibitem [{\citenamefont {Werner}\ \emph {et~al.}(2006)\citenamefont {Werner},
  \citenamefont {Comanac}, \citenamefont {de' Medici}, \citenamefont {Troyer},\
  and\ \citenamefont {Millis}}]{Werner2006}%
  \BibitemOpen
  \bibfield  {author} {\bibinfo {author} {\bibfnamefont {Philipp}\ \bibnamefont
  {Werner}}, \bibinfo {author} {\bibfnamefont {Armin}\ \bibnamefont {Comanac}},
  \bibinfo {author} {\bibfnamefont {Luca}\ \bibnamefont {de' Medici}}, \bibinfo
  {author} {\bibfnamefont {Matthias}\ \bibnamefont {Troyer}}, \ and\ \bibinfo
  {author} {\bibfnamefont {Andrew~J.}\ \bibnamefont {Millis}},\ }\bibfield
  {title} {\enquote {\bibinfo {title} {Continuous-time solver for quantum
  impurity models},}\ }\href {\doibase 10.1103/PhysRevLett.97.076405}
  {\bibfield  {journal} {\bibinfo  {journal} {Phys. Rev. Lett.}\ }\textbf
  {\bibinfo {volume} {97}},\ \bibinfo {pages} {076405} (\bibinfo {year}
  {2006})}\BibitemShut {NoStop}%
\bibitem [{\citenamefont {Parcollet}\ \emph {et~al.}(2015)\citenamefont
  {Parcollet}, \citenamefont {Ferrero}, \citenamefont {Ayral}, \citenamefont
  {Hafermann}, \citenamefont {Krivenko}, \citenamefont {Messio},\ and\
  \citenamefont {Seth}}]{Parcollet2015}%
  \BibitemOpen
  \bibfield  {author} {\bibinfo {author} {\bibfnamefont {Olivier}\ \bibnamefont
  {Parcollet}}, \bibinfo {author} {\bibfnamefont {Michel}\ \bibnamefont
  {Ferrero}}, \bibinfo {author} {\bibfnamefont {Thomas}\ \bibnamefont {Ayral}},
  \bibinfo {author} {\bibfnamefont {Hartmut}\ \bibnamefont {Hafermann}},
  \bibinfo {author} {\bibfnamefont {Igor}\ \bibnamefont {Krivenko}}, \bibinfo
  {author} {\bibfnamefont {Laura}\ \bibnamefont {Messio}}, \ and\ \bibinfo
  {author} {\bibfnamefont {Priyanka}\ \bibnamefont {Seth}},\ }\bibfield
  {title} {\enquote {\bibinfo {title} {Triqs: A toolbox for research on
  interacting quantum systems},}\ }\href {\doibase
  http://dx.doi.org/10.1016/j.cpc.2015.04.023} {\bibfield  {journal} {\bibinfo
  {journal} {Computer Physics Communications}\ }\textbf {\bibinfo {volume}
  {196}},\ \bibinfo {pages} {398 -- 415} (\bibinfo {year} {2015})}\BibitemShut
  {NoStop}%
\bibitem [{\citenamefont {Seth}\ \emph {et~al.}(2016)\citenamefont {Seth},
  \citenamefont {Krivenko}, \citenamefont {Ferrero},\ and\ \citenamefont
  {Parcollet}}]{Seth2016}%
  \BibitemOpen
  \bibfield  {author} {\bibinfo {author} {\bibfnamefont {Priyanka}\
  \bibnamefont {Seth}}, \bibinfo {author} {\bibfnamefont {Igor}\ \bibnamefont
  {Krivenko}}, \bibinfo {author} {\bibfnamefont {Michel}\ \bibnamefont
  {Ferrero}}, \ and\ \bibinfo {author} {\bibfnamefont {Olivier}\ \bibnamefont
  {Parcollet}},\ }\bibfield  {title} {\enquote {\bibinfo {title} {Triqs/cthyb:
  A continuous-time quantum monte carlo hybridisation expansion solver for
  quantum impurity problems},}\ }\href {\doibase
  http://dx.doi.org/10.1016/j.cpc.2015.10.023} {\bibfield  {journal} {\bibinfo
  {journal} {Computer Physics Communications}\ }\textbf {\bibinfo {volume}
  {200}},\ \bibinfo {pages} {274 -- 284} (\bibinfo {year} {2016})}\BibitemShut
  {NoStop}%
\bibitem [{\citenamefont {Troyer}\ and\ \citenamefont
  {Wiese}(2005)}]{Troyer2005}%
  \BibitemOpen
  \bibfield  {author} {\bibinfo {author} {\bibfnamefont {Matthias}\
  \bibnamefont {Troyer}}\ and\ \bibinfo {author} {\bibfnamefont {Uwe-Jens}\
  \bibnamefont {Wiese}},\ }\bibfield  {title} {\enquote {\bibinfo {title}
  {Computational complexity and fundamental limitations to fermionic quantum
  monte carlo simulations},}\ }\href {\doibase 10.1103/PhysRevLett.94.170201}
  {\bibfield  {journal} {\bibinfo  {journal} {Phys. Rev. Lett.}\ }\textbf
  {\bibinfo {volume} {94}},\ \bibinfo {pages} {170201} (\bibinfo {year}
  {2005})}\BibitemShut {NoStop}%
\bibitem [{\citenamefont {Glossop}\ \emph {et~al.}(2011)\citenamefont
  {Glossop}, \citenamefont {Kirchner}, \citenamefont {Pixley},\ and\
  \citenamefont {Si}}]{Glossop2011}%
  \BibitemOpen
  \bibfield  {author} {\bibinfo {author} {\bibfnamefont {Matthew~T.}\
  \bibnamefont {Glossop}}, \bibinfo {author} {\bibfnamefont {Stefan}\
  \bibnamefont {Kirchner}}, \bibinfo {author} {\bibfnamefont {J.~H.}\
  \bibnamefont {Pixley}}, \ and\ \bibinfo {author} {\bibfnamefont {Qimiao}\
  \bibnamefont {Si}},\ }\bibfield  {title} {\enquote {\bibinfo {title}
  {Critical kondo destruction in a pseudogap anderson model: Scaling and
  relaxational dynamics},}\ }\href {\doibase 10.1103/PhysRevLett.107.076404}
  {\bibfield  {journal} {\bibinfo  {journal} {Phys. Rev. Lett.}\ }\textbf
  {\bibinfo {volume} {107}},\ \bibinfo {pages} {076404} (\bibinfo {year}
  {2011})}\BibitemShut {NoStop}%
\bibitem [{\citenamefont {Chowdhury}\ and\ \citenamefont
  {Ingersent}(2015)}]{Chowdhury2015}%
  \BibitemOpen
  \bibfield  {author} {\bibinfo {author} {\bibfnamefont {Tathagata}\
  \bibnamefont {Chowdhury}}\ and\ \bibinfo {author} {\bibfnamefont {Kevin}\
  \bibnamefont {Ingersent}},\ }\bibfield  {title} {\enquote {\bibinfo {title}
  {Critical charge fluctuations in a pseudogap anderson model},}\ }\href
  {\doibase 10.1103/PhysRevB.91.035118} {\bibfield  {journal} {\bibinfo
  {journal} {Phys. Rev. B}\ }\textbf {\bibinfo {volume} {91}},\ \bibinfo
  {pages} {035118} (\bibinfo {year} {2015})}\BibitemShut {NoStop}%
\bibitem [{\citenamefont {Krishna-murthy}\ \emph
  {et~al.}(1980{\natexlab{a}})\citenamefont {Krishna-murthy}, \citenamefont
  {Wilkins},\ and\ \citenamefont {Wilson}}]{Krishna-murthy1980b}%
  \BibitemOpen
  \bibfield  {author} {\bibinfo {author} {\bibfnamefont {H.~R.}\ \bibnamefont
  {Krishna-murthy}}, \bibinfo {author} {\bibfnamefont {J.~W.}\ \bibnamefont
  {Wilkins}}, \ and\ \bibinfo {author} {\bibfnamefont {K.~G.}\ \bibnamefont
  {Wilson}},\ }\bibfield  {title} {\enquote {\bibinfo {title}
  {Renormalization-group approach to the anderson model of dilute magnetic
  alloys. ii. static properties for the asymmetric case},}\ }\href {\doibase
  10.1103/PhysRevB.21.1044} {\bibfield  {journal} {\bibinfo  {journal} {Phys.
  Rev. B}\ }\textbf {\bibinfo {volume} {21}},\ \bibinfo {pages} {1044--1083}
  (\bibinfo {year} {1980}{\natexlab{a}})}\BibitemShut {NoStop}%
\bibitem [{\citenamefont {Horvati\'c}\ and\ \citenamefont
  {Zlati\'c}(1980)}]{Horvatic1980}%
  \BibitemOpen
  \bibfield  {author} {\bibinfo {author} {\bibfnamefont {B.}~\bibnamefont
  {Horvati\'c}}\ and\ \bibinfo {author} {\bibfnamefont {V.}~\bibnamefont
  {Zlati\'c}},\ }\bibfield  {title} {\enquote {\bibinfo {title} {Perturbation
  expansion for the asymmetric anderson hamiltonian},}\ }\href {\doibase
  https://doi.org/10.1002/pssb.2220990125} {\bibfield  {journal} {\bibinfo
  {journal} {Phys. Status Solidi (b)}\ }\textbf {\bibinfo {volume} {99}},\
  \bibinfo {pages} {251--266} (\bibinfo {year} {1980})}\BibitemShut {NoStop}%
\bibitem [{\citenamefont {Bulla}\ \emph {et~al.}(1998)\citenamefont {Bulla},
  \citenamefont {Hewson},\ and\ \citenamefont {Pruschke}}]{Bulla1998}%
  \BibitemOpen
  \bibfield  {author} {\bibinfo {author} {\bibfnamefont {R}~\bibnamefont
  {Bulla}}, \bibinfo {author} {\bibfnamefont {A~C}\ \bibnamefont {Hewson}}, \
  and\ \bibinfo {author} {\bibfnamefont {Th}~\bibnamefont {Pruschke}},\
  }\bibfield  {title} {\enquote {\bibinfo {title} {Numerical renormalization
  group calculations for the self-energy of the impurity anderson model},}\
  }\href {\doibase 10.1088/0953-8984/10/37/021} {\bibfield  {journal} {\bibinfo
   {journal} {Journal of Physics: Condensed Matter}\ }\textbf {\bibinfo
  {volume} {10}},\ \bibinfo {pages} {8365--8380} (\bibinfo {year}
  {1998})}\BibitemShut {NoStop}%
\bibitem [{\citenamefont {Dang}\ \emph
  {et~al.}(2014{\natexlab{a}})\citenamefont {Dang}, \citenamefont {Millis},\
  and\ \citenamefont {Marianetti}}]{Dang2014a}%
  \BibitemOpen
  \bibfield  {author} {\bibinfo {author} {\bibfnamefont {Hung~T.}\ \bibnamefont
  {Dang}}, \bibinfo {author} {\bibfnamefont {Andrew~J.}\ \bibnamefont
  {Millis}}, \ and\ \bibinfo {author} {\bibfnamefont {Chris~A.}\ \bibnamefont
  {Marianetti}},\ }\bibfield  {title} {\enquote {\bibinfo {title} {Covalency
  and the metal-insulator transition in titanate and vanadate perovskites},}\
  }\href {\doibase 10.1103/PhysRevB.89.161113} {\bibfield  {journal} {\bibinfo
  {journal} {Phys. Rev. B}\ }\textbf {\bibinfo {volume} {89}},\ \bibinfo
  {pages} {161113} (\bibinfo {year} {2014}{\natexlab{a}})}\BibitemShut
  {NoStop}%
\bibitem [{\citenamefont {Dang}\ \emph
  {et~al.}(2014{\natexlab{b}})\citenamefont {Dang}, \citenamefont {Millis},\
  and\ \citenamefont {Marianetti}}]{Dang2014b}%
  \BibitemOpen
  \bibfield  {author} {\bibinfo {author} {\bibfnamefont {Hung~T.}\ \bibnamefont
  {Dang}}, \bibinfo {author} {\bibfnamefont {Andrew~J.}\ \bibnamefont
  {Millis}}, \ and\ \bibinfo {author} {\bibfnamefont {Chris~A.}\ \bibnamefont
  {Marianetti}},\ }\bibfield  {title} {\enquote {\bibinfo {title} {Covalency
  and the metal-insulator transition in titanate and vanadate perovskites},}\
  }\href {\doibase 10.1103/PhysRevB.89.161113} {\bibfield  {journal} {\bibinfo
  {journal} {Phys. Rev. B}\ }\textbf {\bibinfo {volume} {89}},\ \bibinfo
  {pages} {161113} (\bibinfo {year} {2014}{\natexlab{b}})}\BibitemShut
  {NoStop}%
\bibitem [{\citenamefont {Dai}\ \emph {et~al.}(2005)\citenamefont {Dai},
  \citenamefont {Haule},\ and\ \citenamefont {Kotliar}}]{Dai2005}%
  \BibitemOpen
  \bibfield  {author} {\bibinfo {author} {\bibfnamefont {Xi}~\bibnamefont
  {Dai}}, \bibinfo {author} {\bibfnamefont {Kristjan}\ \bibnamefont {Haule}}, \
  and\ \bibinfo {author} {\bibfnamefont {Gabriel}\ \bibnamefont {Kotliar}},\
  }\bibfield  {title} {\enquote {\bibinfo {title} {Strong-coupling solver for
  the quantum impurity model},}\ }\href {\doibase 10.1103/PhysRevB.72.045111}
  {\bibfield  {journal} {\bibinfo  {journal} {Phys. Rev. B}\ }\textbf {\bibinfo
  {volume} {72}},\ \bibinfo {pages} {045111} (\bibinfo {year}
  {2005})}\BibitemShut {NoStop}%
\bibitem [{\citenamefont {Yosida}\ and\ \citenamefont
  {Yamada}(1975)}]{Yosida1975}%
  \BibitemOpen
  \bibfield  {author} {\bibinfo {author} {\bibfnamefont {Kei}\ \bibnamefont
  {Yosida}}\ and\ \bibinfo {author} {\bibfnamefont {Kosaku}\ \bibnamefont
  {Yamada}},\ }\bibfield  {title} {\enquote {\bibinfo {title} {{Perturbation
  Expansion for the Anderson Hamiltonian. III}},}\ }\href {\doibase
  10.1143/PTP.53.1286} {\bibfield  {journal} {\bibinfo  {journal} {Progress of
  Theoretical Physics}\ }\textbf {\bibinfo {volume} {53}},\ \bibinfo {pages}
  {1286--1301} (\bibinfo {year} {1975})}\BibitemShut {NoStop}%
\bibitem [{\citenamefont {Tong}(2015)}]{Tong2015}%
  \BibitemOpen
  \bibfield  {author} {\bibinfo {author} {\bibfnamefont {Ning-Hua}\
  \bibnamefont {Tong}},\ }\bibfield  {title} {\enquote {\bibinfo {title}
  {Equation-of-motion series expansion of double-time green's functions},}\
  }\href {\doibase 10.1103/PhysRevB.92.165126} {\bibfield  {journal} {\bibinfo
  {journal} {Phys. Rev. B}\ }\textbf {\bibinfo {volume} {92}},\ \bibinfo
  {pages} {165126} (\bibinfo {year} {2015})}\BibitemShut {NoStop}%
\bibitem [{\citenamefont {El-Batanouny}(2020)}]{El-Batanouny2020}%
  \BibitemOpen
  \bibfield  {author} {\bibinfo {author} {\bibfnamefont {Michael}\ \bibnamefont
  {El-Batanouny}},\ }\href@noop {} {\emph {\bibinfo {title} {Advanced Quantum
  Condensed Matter Physics: One-Body, Many-Body, and Topological
  Perspectives}}}\ (\bibinfo  {publisher} {Cambridge University Press},\
  \bibinfo {year} {2020})\BibitemShut {NoStop}%
\bibitem [{\citenamefont {Franz}()}]{Franz2019}%
  \BibitemOpen
  \bibfield  {author} {\bibinfo {author} {\bibfnamefont {Marcel}\ \bibnamefont
  {Franz}},\ }\href {https://phas.ubc.ca/~franz/course/cm525/} {\enquote
  {\bibinfo {title} {Advanced condensed matter physics: Topological insulators
  and superconductors},}\ }\bibinfo {note} {Course Materials}\BibitemShut
  {NoStop}%
\bibitem [{\citenamefont {Krishna-murthy}\ \emph
  {et~al.}(1980{\natexlab{b}})\citenamefont {Krishna-murthy}, \citenamefont
  {Wilkins},\ and\ \citenamefont {Wilson}}]{Krishna-murthy1980a}%
  \BibitemOpen
  \bibfield  {author} {\bibinfo {author} {\bibfnamefont {H.~R.}\ \bibnamefont
  {Krishna-murthy}}, \bibinfo {author} {\bibfnamefont {J.~W.}\ \bibnamefont
  {Wilkins}}, \ and\ \bibinfo {author} {\bibfnamefont {K.~G.}\ \bibnamefont
  {Wilson}},\ }\bibfield  {title} {\enquote {\bibinfo {title}
  {Renormalization-group approach to the anderson model of dilute magnetic
  alloys. i. static properties for the symmetric case},}\ }\href {\doibase
  10.1103/PhysRevB.21.1003} {\bibfield  {journal} {\bibinfo  {journal} {Phys.
  Rev. B}\ }\textbf {\bibinfo {volume} {21}},\ \bibinfo {pages} {1003--1043}
  (\bibinfo {year} {1980}{\natexlab{b}})}\BibitemShut {NoStop}%
\end{thebibliography}%

\end{document}